\documentclass[a4paper,11pt]{article}

\usepackage{amssymb}
\usepackage{graphicx} 
\usepackage{times}
\usepackage{fullpage}
\usepackage{color}
\usepackage{enumerate}
\usepackage{amsthm}

\newcommand{\smallcycle}{$\frac{\log n}{10\log d}\:$}

\newcommand{\remove}[1]{}

\newtheorem{myclaim}{Claim}[section]

\newenvironment{myproof}{\noindent{\itshape Proof\@:}}{\hfill $\Diamond$\\}
 
\newtheorem{definition}{Definition}[section]

\newtheorem{theorem}{Theorem}[section]

\newtheorem{lemma}{Lemma}[section]

\newtheorem{proposition}{Proposition}[section]

\newtheorem{corollary}{Corollary}[section]

\newenvironment{propositionproof}[1]{\noindent{\bf Proof of Proposition #1\@:}}{\hfill $\Diamond$\\}

\date{\today}
\title{Deterministic Counting of Graph Colourings\\ Using Sequences of Subgraphs}
\author{Charilaos Efthymiou\\ Goethe University, Mathematics Institute, Frankfurt 60054, Germarny\\
{\tt efthymiou@gmail.com}}

\begin{document}

\maketitle

\begin{abstract}
In this paper we propose a deterministic algorithm for approximately counting the $k$-colourings of sparse
random graphs $G(n,d/n)$.  In particular, our algorithm computes in polynomial time a $(1\pm n^{-\Omega(1)})$-ap\-pro\-xi\-ma\-tion
of the logarithm of the number of $k$-colourings of $G(n,d/n)$ for $k\geq (2+\epsilon) d$ with high
probability over the graph instances.

Our algorithm is related to the algorithms of A.~ Bandyopadhyay et al. in SODA '06, and A.~Montanari et al.
in SODA '06,  i.e. it uses {\em spatial correlation decay} to compute {\em deterministically} marginals of
{\em Gibbs distribution}. We develop a scheme whose accuracy depends on {\em non-reconstruction} of the
colourings of $G(n,d/n)$, rather than {\em uniqueness} that are required in previous works.
This leaves open the possibility for our schema to be sufficiently accurate even for $k<d$.

The set up for establishing correlation decay is as follows: Given $G(n,d/n)$,  we alter the graph  structure
in some specific region $\Lambda$ of the graph by deleting edges between vertices of $\Lambda$. Then we show 
that the effect of this change on the marginals of Gibbs distribution, diminishes as we move away from $\Lambda$.
Our approach is novel and suggests a new context for the study of deterministic counting algorithms.
\end{abstract}

\section{Introduction}
For a graph $G=(V,E)$ and a positive integer $k$, a proper $k$-colouring is
an assignment $\sigma:V\to [k]$ (we use $[k]$ to denote $\{1,\dots, k \}$),
such that adjacent vertices receive different members of $[k]$,  i.e. 
different ``colours''. Here we focus on the problem of {\em counting} the
$k$-colourings of $G$.  In particular, we consider the cases where the
underlying graph is an instance of Erd\H{o}s-R\'enyi random graph $G(n,p)$,
where $p=d/n$ and $d$ is `large' but remains bounded as $n\to \infty$.
We say that an event occurs {\em with high probability (w.h.p.)} if the
probability of the event to occur tends to 1 as $n\to \infty$.

Usually, a counting problem is reduced to computing marginal probabilities
of {\em Gibbs distribution}, see \cite{SelfReduc}. Typically, we estimate
these marginals by using a {\em sampling} algorithm.  The most powerful
method for sampling is the Markov Chain Monte Carlo (MCMC). There the main
technical challenge is to establish that the underlying Markov chain mixes
in polynomial time (see \cite{jerrum,countD}). The MCMC method gives {\em probabilistic}
approximation guarantees.

Recently, new approaches were proposed for {\em deterministic} counting algorithms
in \cite{long-partition-guide} and \cite{weitz-alltrees}. 
The work in \cite{long-partition-guide} is for counting colourings and independent 
sets, while \cite{weitz-alltrees} is for independent sets.
These new approaches link the correlation decay to {\em computing efficiently} marginals of Gibbs distributions.
The two algorithms in \cite{long-partition-guide,weitz-alltrees} suggest two 
different approaches for computing marginals.  The one in \cite{long-partition-guide}
applies mainly to locally tree graphs. Spatial correlation decay is exploited so as to
restrict the computations of marginals and  consider only small areas of the graph.
The accuracy of the computations there relies on establishing the so-called {\em uniqueness 
conditions} on trees. On the other hand, the algorithm in \cite{weitz-alltrees} applies
to a wider family of graphs, i.e. not necessarily locally treelike ones. It uses a more
elaborate technique which somehow handles the existence of cycles in the computation 
of marginals, mainly by fixing the spins of certain sites appropriately.  The approximation
guarantees for the second algorithm are stronger than those of the first one. However, the
stronger results do not come for free. The spatial mixing assumptions there are stronger,
e.g. for the case of counting independent sets it requires {\em strong spatial mixing
conditions}.

Our approach for computing Gibbs marginals is closer to \cite{long-partition-guide} as 
w.h.p. the instance of $G(n,d/n)$ is locally tree like. However, this is not just an 
extension of \cite{long-partition-guide} to random graphs. 
First we express the bounds for $k$ in terms of the expected degree of the graph,
rather than the maximum degree which is the case in \cite{long-partition-guide}.
Furthermore, we relate the computation of Gibbs marginals to weaker notions of 
spatial mixing, namely the so-called {\em non
reconstruction conditions}. Compared to Gibbs uniqueness condition, which is required
in \cite{long-partition-guide}, non-reconstruction is weaker and holds for a wider 
range of $k$. This leaves open the possibility for our schema to be sufficiently 
accurate for counting $k$-colourings of $G(n,d/n)$ even for $k<d$, i.e. when uniqueness 
condition is not expected to hold.

\paragraph{Further Motivation.}
Apart from its use for counting algorithms, the problem of computing efficiently good 
approximations of Gibbs marginals is a very interesting problem on its own. It is related
to the empirical success of heuristics suggested by statistical physicists such as
{\em Belief Propagation} and {\em Survey Propagation} (see e.g.  \cite{lenka}). In 
theoretical computer science, these heuristics are studied in the context of finding
solution of random instances of Constraint Satisfaction Problems, e.g. random graph 
colouring, random $k$-SAT, etc. Similar ideas for computing marginals were also suggested
in {\em coding theory} and {\em artificial intelligence} (see in \cite{InfPhyComp}).

\paragraph{Related Work.}
Algorithms that follow  a similar approach  as the one in \cite{long-partition-guide}, appear in
\cite{KSAT,TCS-sampling}. The one in \cite{KSAT} is for computing  Gibbs marginals for random 
instances of $k$-SAT. The one in \cite{TCS-sampling} is for random colouring of $G(n,d/n)$. The 
algorithm in \cite{TCS-sampling} does not compute the log partition function, however, it can be 
altered so as to do so. Then, it is not hard to show that it requires at least $d^{7/2}$ colours.

On the other hand, counting algorithms as the one in \cite{weitz-alltrees} give better polynomial time 
approximations, compared to the ones referred in the previous paragraph. However, 
they require stronger correlation decay conditions. Attempts to establish such strong
conditions were successful for two spin cases, e.g. independent sets, matchings, 
Ising spins (see \cite{weitz-alltrees,FPTAS-matchings,mossel-ising-gnp}]). For the 
multi-spin cases, such as colourings, things seem harder. The best algorithm of
this category for counting k-colourings requires $k > 2.8\Delta$ and girth at least 4
(see \cite{FPTAS}), where $\Delta$ is the maximum degree of the underlying graph.

The author of this work, in a subsequent paper \cite{MySODA12}, uses some of the ideas that
appear here in an algorithm for approximate random colouring $G(n,d/n)$. The algorithm 
there yields similar results as here but the approximation guarantees are probabilistic ones,
i.e. the same as the Monte Carlo algorithms.

\subsection{Results}\label{sec:FormResults}
Let $Z(G,k)$ denote the number of $k$-colourings of the graph $G$. In statistic physics literature the
quantity $Z(G,k)$ is also known as the {\em partition function}. Our algorithm computes an approximation
for the log-partition function $\log Z(G,k)$.

\begin{definition}\label{def:FreeEnergyApprox}
$\Psi$ is defined to be an $\epsilon$-approximation of the log-partition function $\log Z(G,k)$ if
\begin{displaymath}
(1-\epsilon)\frac{\log Z(G,k)}{n}\leq \Psi\leq
(1+\epsilon)\frac{\log Z(G,k)}{n}.
\end{displaymath}
\end{definition}

\noindent
The results of our work are the following ones:

\begin{theorem}\label{thrm:GnpApp}
Let $\epsilon>0$ be a fixed number and let $d$ be sufficiently large. For $k\geq (2+\epsilon)d$ and 
with probability at least $1-n^{-a}$, over the graph instances,  our algorithm {computes} an $n^{-b}$-approximation 
of  $\log Z(G_{n,d/n}, k)$, in time $O(n^{s})$, where  $a$, $b$ and $s$ are positive real numbers 
which depend on $k$.
\end{theorem}

\noindent  
Roughly speaking the above theorem implies that for typical instances of $G(n,d/n)$ and $k\geq (2+\epsilon)d$ our 
algorithm is able to compute Gibbs marginals of the $k$-colourings of $G(n,d/n)$ within error $o(n^{-c})$, where 
$c>0$ is fixed.  Furthermore, the fact that the Gibbs distribution of $k$-colourings is symmetric and the fact 
that w.h.p. all but a vanishing fraction of the edges in $G(n,d/n)$ do not belong to cycles shorter than 
$\Theta(\ln n)$ implies the following result.

\begin{corollary}\label{theorem:free-nrg}
For sufficiently large $d$  and $k\geq (2+\epsilon) d$, w.h.p. it holds that 
\begin{displaymath}
\left |\frac{\log Z(G_{n,d/n},k)}{n}- \left( \log k+ \frac{d}{2} \cdot
\log\left( 1-\frac{1}{k}\right )\right) \right| \leq {n^{-c}},
\end{displaymath}
for fixed $c>0$.
\end{corollary}

\noindent
Observe that the concentration result in Corollary \ref{theorem:free-nrg}, for the number of $k$-colouring
of $G(n,d/n)$, is derived by using correlation decay arguments.  In the literature of random structures
such results are typically derived by using the so-called ``Second moment method''. A less accurate result
can be derived from the work of  Achlioptas and Naor in \cite{optas-chrno} with some extra work,  i.e.
the error there is $O(\log^{-1} n)$.

Finally, a related question and somehow a natural one is whether we can distinguish efficiently the instances
of $G(n,d/n)$ that have their log-partition function concentrated. That is, for a sufficiently large function
$h(n,d,k)$ we can answer whether a given instance $G(n,d/n)$ is such that 
$$
\left |\frac{\log Z(G(n,d/n),k)}{n}- \left( \log k+ \frac{d}{2} \cdot \log\left( 1-\frac{1}{k}\right )\right) \right| \leq h(n,k,d),
$$
or not.  This goes beyond what we can get from the second  moment method, as the later uses non-constructive arguments.
We show that such distinction of instances is possible. The reason is that our arguments for correlation decay are tightly 
related to the degrees of vertices. That is, {\em examining the degrees of the vertices} we can infer whether the number
of colourings of $G(n,d/n)$ is concentrated.

Let $S(n,d)$ denote the set of graphs on $n$ vertices which have the following
properties: Their number of edges is at most $3dn/4$. There are at most $n^{0.3}$
cycles, each of them, of length at most $\frac{\log n}{10\log d}$.
Finally, for each vertex $v$ in the graph, the induced subgraph that contains
$v$ and all vertices within distance $\frac{\log n}{4\log(e^2d/2)}$ is either
tree or a unicyclic graph. In the following result, we show that for the graphs 
in $S(n,d/n)$ it is possible to verify whether the log-partition function is
concentrated or not.

\begin{corollary}\label{thrm:verification}
Let $\epsilon>0$ be a fixed number and let $d$ be sufficiently large.
For $k\geq (2+\epsilon) d$, there exists a set of graphs $S(n,d)$ 
such that the following holds: 
For any sufficiently large real function $h(n,d,k)\geq n^{-O(1)}$ 
it  can be verified in polynomial time whether the property
\begin{eqnarray}\label{eq:VerifCond}
\left |\frac{\log Z(G,k)}{n}- \left( \log k+ \frac{d}{2} \cdot \log\left( 1-\frac{1}{k}\right )\right) \right| \leq h(n,k,d).
\end{eqnarray}
holds or not, for any $G\in S(n,d)$. Furthermore, $Pr[G_{n,d/n}\in S(n,d)]=1-n^{-0.1}$ 
and deciding whether $G_{n,d/n}\in S(n,d)$ can be made in polynomial time.
\end{corollary}


\subsection{Contribution}\label{sec:contribution}
We could partition the contribution of our work into two parts. The first part includes 
a new approximation-schema for computing deterministically Gibbs marginals. In the second 
part we present the tool for bounding correlation decay quantities that arise in the schema.
\\ \vspace{-.3cm}

\noindent
{\bf Approximating Gibbs Marginals.}
The problem of counting $k$-colourings of a graph $G=(V,E)$ reduces to the problem of estimating Gibbs marginals which can be formulated as follows:
\\ \vspace{-.3cm}

\noindent
{\em Problem 1.} Consider the graph $G=(V,E)$ and let $\mu(\cdot)$ denote the Gibbs distribution
over the proper $k$-colourings of $G$. For the small (fixed sized) set of vertices $\Lambda \subset V$
and for $\sigma_{\Lambda}\in [k]^{\Lambda}$, compute the probability $\mu(\sigma_{\Lambda})$.
\\ \vspace{-.3cm}

\noindent
In the general case computing $\mu(\sigma_{\Lambda})$ exactly  requires superpolynomial time. So
the focus is on approximating it. One possible approach for computing an approximation of the
marginal in Problem 1 was suggested in \cite{long-partition-guide} for locally tree graphs.
Roughly speaking the idea can be described as follows: The Gibbs marginal on $\Lambda$ can be
expressed as a convex combination of boundary conditions on $L_{t, \Lambda}$, the vertices at
distance $t$  from $\Lambda$, as follows 

\begin{equation}\label{eq:convexcombination}
\mu(\sigma_{\Lambda})=\sum_{\tau \in [k]^{L_{t, \Lambda}}}\mu(\sigma_{\Lambda}|\tau)\mu(\tau).
\end{equation}
Pick $t$ such that we can compute in polynomial time each of the marginals $\mu(\sigma_{\Lambda}|\tau)$. 
The problem, then, reduces to the not easier task of computing the coefficients  $\mu(\tau)$.
The authors  in \cite{long-partition-guide} noticed that the problem of estimating these
coefficients somehow ``degenerates'' if $k$ is so large that the marginals $\mu(\sigma_{\Lambda}|\tau)$
and $\mu(\sigma_{\Lambda}|\tau')$  are sufficiently close to each other, for any $\tau, \tau' \in [k]^{L_{t, \Lambda}}$
in the support of $\mu$. In this case, the convexity implies that $\mu(\sigma_{\Lambda})$
is sufficiently close to any of the conditional marginals in the r.h.s. of (\ref{eq:convexcombination}).
Using this observation and the fact that we have chosen $t$ such that the conditional marginals can be
computed in polynomial time, it is direct that the above schema gives in polynomial time an
approximation of $\mu(\sigma_{\Lambda})$.

We should remark that the conditional marginals above are close to each other if a certain kind 
of independence hold, between the colourings of $\Lambda$ and the colourings
of $L_{t, \Lambda}$. Establishing such a kind of independence  is related to  what is known
in statistical physics as establishing  ``Dobrushin Uniqueness Condition'' (see \cite{BiblosSpin}).

Our approach, here, is in a similar spirit. However, it amounts to substituting the coefficients
$\mu(\tau)$ with new, different, ones. The aim is not to bypass the estimation of coefficients but
somehow to {\em approximate} them. So instead of $G$ we consider the graph $G_{t, \Lambda}$, the
induced subgraph of $G$ that contains the set $\Lambda$ and all its neighbours within graph distance
$t$.  We denote with $\hat{\mu}(\sigma_{\Lambda})$ the new Gibbs marginal of the event 
$\sigma_{\Lambda}$  in the $k$-colourings of $G_{t, \Lambda}$. We will use $\hat{\mu}(\sigma_{\Lambda})$
to approximate $\mu(\sigma_{\Lambda})$. Note that we have chosen $t$ so as the computation of
$\hat{\mu}(\sigma_{\Lambda})$ can be carried out efficiently. Writing the corresponding of 
(\ref{eq:convexcombination}) for the graph $G_{t, \Lambda}$ we get that
$$
\hat{\mu}(\sigma_{\Lambda})=\sum_{\tau\in [k]^{L_t(\Lambda)}}\hat{\mu}(\sigma_{\Lambda}|\tau)\hat{\mu}(\tau).
$$

\noindent
{\em Remark 1.}
Someone could use uniqueness condition here as well, i.e. work as in 
\cite{long-partition-guide}. However, here we make a more detailed comparison 
of $\hat{\mu}(\sigma_{\Lambda})$ and ${\mu}(\sigma_{\Lambda})$. As a matter of
fact, our analysis gives rise to non-reconstruction spatial mixing conditions.
\\ \vspace{-.3cm}

\noindent
The key observation to compare $\hat{\mu}$ and $\mu$ is the following one: The distribution
$\hat{\mu}(\cdot)$ can be seen as being induced by the deletion of the edges that connect the
neighbourhood $G_{t, \Lambda}$  with the rest of the graph $G$. We require that the deletion
of these edges does not have great effect on the marginals on $\Lambda$. 
It turns out that this is equivalent to requiring {\em non-reconstructibility}
condition \footnote{Non-reconstructibility is equivalent to extremality of Gibbs measure for
infinite graphs, see e.g. \cite{BiblosSpin}.} with (sufficiently fast) exponential decay. That
is, let  $G'$ be either  $G$ (the graph in Problem 1) or any of its subgraph. Let $\mu'$ be the
Gibbs distribution of the colourings of $G'$. Then, {\em non-reconstructibility} condition with
exponential decay can be expressed as follows: 

\begin{equation}\label{eq:SingleSiteBoundary}
\max_{{\cal C}\in [k]^x} ||\mu'(\cdot)-\mu'(\cdot |{\cal C})||_{L_{x,t}}\leq \exp(-at),
\end{equation}
where $x$ is a vertex in $G'$, $L_{t,x}$ contains all the vertices which are at distance $t$ from $x$
and $\alpha>0$ is a fixed number.

For the distributions $\nu_{a}, \nu_{b}$ on $[k]^V$, we let $|| \nu_{a}-\nu_{b}||$ denote their
{\em total variation distance}, i.e.
\begin{equation}\label{eq:TVDDefinition}
|| \nu_{a}-\nu_{b}||=\max_{\Omega' \subseteq [k]^V} | \nu_{a}(\Omega')-\nu_{b}(\Omega')|.
\end{equation}
For $\Lambda \subseteq V$ let $||\nu_{a}-\nu_{b}||_{\Lambda}$ denote the total variation distance between the projections of  $\nu_a$ and $\nu_{b}$ on $[k]^{\Lambda}$.

\paragraph{Bounds for Spatial Correlation Decay.} 
We complement the new approach for estimating Gibbs marginals, by providing a general 
tool for bounding  correlation decay conditions as in (\ref{eq:SingleSiteBoundary}).
We bound the correlation between some vertex $x$ and the vertices at distance $t$ from
$x$ by studying the probability of the following event: Choose u.a.r. a $k$-colouring 
of $G'$. Let $\rho$ be the probability that there are two colour classes that specify
a {\em connected} subgraph of $G'$ that contains both $x$ and some vertices at distance
$t$. Then we show that 
$\max_{{\cal C}\in [k]^x} ||\mu'(\cdot)-\mu'(\cdot|{\cal C})||_{L_{x,t}}\leq \rho$.

We derive bounds for the quantity $\rho$ by using the well-known technique from statistical
physics called ``disagreement percolation'' coupling construction \cite{disagreement-percolation}.
It turns out that using the disagreement percolation we express the decay of  correlation
as in (\ref{eq:SingleSiteBoundary}) in terms of percolation-probabilities on the graph. 
Our technique is general and simple, e.g. there is no need for restrictions on the graph
structure which was the case in \cite{long-partition-guide,TCS-sampling,KSAT}.  Furthermore, 
it allows expressing the corresponding  bounds  in terms of the degree of each vertex, not the maximum degree.
\\ \vspace{-.2cm}

\noindent
{\em Remark 2.}
``Disagreement Percolation'' has been used for bounding different kinds of correlation decay in 
works for MCMC sampling colouring, e.g. \cite{MCRandColPlanar,old-GnpSampling}. Also, disagreement
percolation appears (implicitly) in \cite{tree-NonReconNaja} as part of a more general technique
for showing non-reconstruction for colourings on trees. Our setting here is more general than
\cite{tree-NonReconNaja} as it considers graphs with cycles. i.e. there are technical issues 
that need to be addressed. 
\\ \vspace{-.2cm}

\noindent
{\em Remark 3.} 
For the sparse random graphs with bounded expected degree $d$ there is a work by Montanari et al. 
in \cite{ReconstructionMontanaryTetali} that shows non-reconstruc\-ti\-bi\-li\-ty for $k$ smaller
than  what we derive here. Unfortunately, we cannot use this result here, mainly, because it does
not imply that the corresponding spatial mixing conditions are monotone in the graph structure.
Note that if we could use the non-reconstructibility bounds from \cite{ReconstructionMontanaryTetali}, 
then our results for counting would be even better.

\subsection{Structure of the paper}
The rest of the paper is organized as follows: In Section \ref{sec:basics} we present some basic
concepts and describe the counting to marginal estimation reduction. In Section \ref{sec:Schema}
we give a general description of our counting algorithm and relate its accuracy with certain kind 
of spatial correlation decay conditions.  Then, we provide the results which are used 
for bounding spatial correlation decay (in Section \ref{sec:StmntSMBounds}).

In Section \ref{sec:GnpAppMain} we discuss the technical details for applying the counting algorithm  
on $G_{n,d/n}$. We prove Theorem \ref{thrm:GnpApp}, Corollary \ref{theorem:free-nrg} and Corollary 
\ref{thrm:verification}. In Section \ref{sec:DisPercBounds} we prove the results that appear in 
Section \ref{sec:StmntSMBounds}, for bounding spatial correlation decay. 
Finally, in Section \ref{sec:RestProof} we provide the proofs of some technical results we use.

\section{Basics and Problem Formulation}\label{sec:basics}

Our algorithm is studied in the context of finite spin-systems, a concept that originates in statistical
physics. In particular, we use the finite {\em colouring model}.
\\ \vspace{-.3cm}
	
\noindent
The {\bf Finite Colouring Model} with underlying graph $G=(V,E)$ that uses $k$ colours is specified by a set of 
``{\em sites}'', which correspond to the vertices of $G$, a set of ``{\em spins}'', i.e.  the set $[k]$, and
a symmetric function $U:[k]\times[k]\to \{0,1\}$ such that for $i,j \in [k]$

\begin{displaymath}
U(i,j )=\left \{
\begin{array}{lcl}
1 &\quad &\textrm{if $i\neq j$}
\\ 
0 &\quad &\textrm{otherwise.}
\end{array}
\right.
\end{displaymath}
We always assume that $k$ is such that $Z(G,k)\neq \emptyset$.

A {\em configuration} $\sigma \in [k]^V$ of the system assigns each vertex (``site'') $x \in V$
the colour (``spin value'') $\sigma_x \in [k]$. The probability to find the system in 
configuration $\sigma$ is determined by the {\em Gibbs distribution}, which is defined as
\begin{displaymath}
\mu(\sigma)=\frac{\prod_{\{x,y\}\in E}U(\sigma_x, \sigma_y)}{Z(G, k)}.
\end{displaymath}

\noindent
It is direct that the Gibbs distribution corresponds to the uniform  distribution over the set 
of $k$-colouring of the underlying graph $G$. A {\em boundary condition} corresponds to fixing 
the colour assignment of a specific ``{\em boundary}'' vertex set of $G$.

Another concept we will need is that of the {\em sequence of subgraphs}.

\begin{definition}[Sequence of subgraphs]\label{definition:subgraph-seq}
For the graph $G=(V,E)$, let ${\cal G}(G)=\{G_i=(V,E_i)\}_{i=0}^r$ denote a sequence of subgraphs  of $G$ which has the following properties:
\begin{itemize}	
	\item $G_{0}$ is  a spanning subgraph of $G$
	\item $E_i \subset E_{i+1}$ for $0\leq i <r $ and $E_r=E$
	\item the term $G_{i+1}$ compared to $G_i$ has 	an additional 
	edge, the edge $\Psi_i=\{v_i,u_i\}$.
\end{itemize}
\end{definition}

\begin{figure}
\centering
\begin{minipage}{0.27\textwidth}
	\centering
		\includegraphics[width=0.8\textwidth]{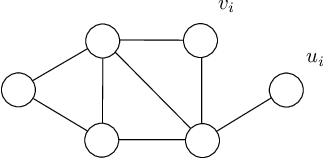}
		\caption{Graph $G_{i}$.}
	\label{fig:GiNew}
\end{minipage}
\begin{minipage}{0.27\textwidth}
	\centering
		\includegraphics[width=0.8\textwidth]{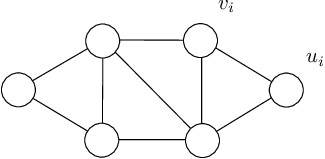}
		\caption{Graph $G_{i+1}$. }
	\label{fig:Gi+1New}
\end{minipage}
\end{figure}

\noindent
When we refer to ${\cal G}(G)$ we specify the graph $G_0$ while we, usually, assume that there is some arbitrary 
rule which gives  the terms $G_1,\ldots, G_r$. In Figures \ref{fig:GiNew} and \ref{fig:Gi+1New} there is an example
of two consecutive terms of a sequence ${\cal G}(G)$,  for some graph $G$. Observe that in $G_i$ the vertices 
$v_i$ and $u_i$ are not adjacent, while in $G_{i+1}$ we add the edge $\Psi_i=\{v_i, u_i\}$.

\begin{lemma}\label{lemma:Count2Sample}
For the graph $G=(V,E)$ consider a sequence of subgraphs ${\cal G}(G)$ where $G_0$ is edgeless. Let $X_i$ 
be a random colouring of $G_i\in{\cal G}(G)$. For some integer $k>0$,  we have that 
$$
|Z(G,k)|=k^n \cdot \prod_{i=1}^{|E|-1}Pr[X_{i}(v_i)\neq X_{i}(u_i)],
$$
where the vertices $v_i$ and $u_i$ are incident to $\Psi_i$. 
\end{lemma}

\noindent
The proof of the above lemma is standard and can be found in various places (e.g. \cite{SelfReduc,countB,countC}), 
for completeness we present it in Section \ref{sec:lemma:Count2Sample}.

We close this section with some additional notation. For $\Lambda \subseteq V$ and some integer $t>0$, we let
$L(\Lambda, t)$ denote the set of vertices at graph distance {\em exactly} $t$ from $\Lambda$. Also, we let 
$B(\Lambda, t)$ denote the set of vertices {\em within} graph distance $t$ from $\Lambda$.

\section{Counting Schema}\label{sec:Schema}

For clarity reasons, we present the counting schema by assuming that
we are given a fixed graph $G=(V,E)$ and some integer $k$ such that $Z(G,k)>0$.

The schema is based on computing Gibbs marginals as it is described in Lemma
\ref{lemma:Count2Sample}. That is, given $G$, we consider a sequence of subgraphs
${\cal G}(G)=G_0, \ldots, G_r$ with $G_0$ being edgeless. For each $G_i\in {\cal G}(G)$
let $X_i$ be a random colouring. In our schema we compute an {\em approximation}
of  each probability term $Pr[X_i(v_i)\neq X_i(u_i)]$ by working as
follows: We consider a new sequence of subgraphs ${\cal G} (G_i)=G_{i,0}, \ldots, G_{i,r_i}$
defined as follows: $G_{i,r_{i}}$, is  the graph $G_i$ while $G_{i, 0}$ is derived
from $G_i$ by removing all the edges between the sets $L(\Psi_i,t)$ and 
$L(\Psi_i, t+1)$\footnote{Both $L(\Psi_i,t)$ and $L(\Psi_i, t+1)$ are considered
w.r.t. graph $G_i$.}, where $t>0$ is some appropriate integer.  We consider $Y_i$
a random colouring of the graph $G_{i,0}\in {\cal G}(G_i)$.  Our schema approximates
$Pr[X(v_i)\neq X(u_i)]$  with $Pr[Y_i(v_i)\neq Y_i(u_i)]$. 

Observe that the computation of $Pr[Y_i(v_i)\neq Y_i(u_i)]$ depends on the 
induced subgraph of $G_i$ which contains only vertices within graph distance $t$ from
$\Psi_i=\{v_i,u_i\}$. Taking sufficiently small $t$ it makes it possible to compute
$Pr[Y_i(v_i)\neq Y_i(u_i)]$ in polynomial time.

\begin{figure}
\centering
\begin{minipage}{0.25\textwidth}
	\centering
		\includegraphics[width=\textwidth]{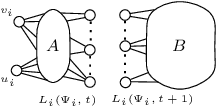}
		\caption{Graph $G_{i0}$.}
	\label{fig:Gir0}
\end{minipage}
\begin{minipage}{0.25\textwidth}
	\centering
		\includegraphics[width=\textwidth]{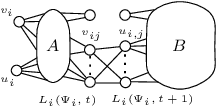}
		\caption{Graph $G_{i,j+1}$.}
	\label{fig:Gij}
\end{minipage}
\begin{minipage}{0.25\textwidth}
	\centering
		\includegraphics[width=\textwidth]{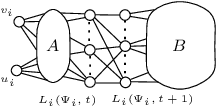}
		\caption{Graph $G_{i,r_i}$.}
	\label{fig:Gi,ri}
\end{minipage}
\end{figure}
Figures \ref{fig:Gir0}, \ref{fig:Gij} and  \ref{fig:Gi,ri} illustrate  some members
of ${\cal G}(G_i)$. That is, Figure \ref{fig:Gir0} shows the first term of the sequence.
Figure \ref{fig:Gij} shows the graph $G_{i,j+1}$, i.e. the edge $\Psi_{i,j}=\{u_{i,j}, v_{u,i}\}$
has just been inserted. In Figure \ref{fig:Gi,ri} we have the final term of 
${\cal G}(G_i)$, the graph $G_{i,r_i}$.

In what follows we provide the pseudocode of the counting algorithm. \\ \vspace{-.3cm}

\noindent
{\bf Counting Schema}\\ \vspace{-.82cm} \\
\rule{\textwidth}{1pt} \\ 
{\em Input}: $G$, $k$, $t$.\\ 
\hspace*{0.3cm}Set ${\cal Z}=k^n$.\\
\hspace*{0.3cm}Compute ${\cal G}(G)=\{G_0,\ldots, G_r\}$.\\
\hspace*{0.3cm}For $0\leq i \leq r-1$ do 
\begin{itemize}
\item Compute ${\cal G}(G_i)$.
\item Compute the exact value of $Pr[Y_{i}(v_i)\neq Y_{i}(u_i)]$.
\item Set ${\cal Z}={\cal Z}\cdot Pr[Y_{i}(v_i)\neq Y_{i}(u_i)]$.
\end{itemize}
\hspace*{0.3cm}End For.\\
{\em Output}: $ \log \left ({\cal Z}\right) /n$.\\ \vspace{-.745cm}\\
\rule{\textwidth}{1pt} \\

\noindent
Two natural questions arise for the counting algorithm. The first one is its 
{\em accuracy}, i.e. how close $\frac{1}{n}\log {\cal Z}$ and $\frac{1}{n}\log Z(G,k)$ are.
The second one is about the {\em time complexity}.

As far as the time complexity is regarded, typically, the execution time is dominated
by the computations for $Pr[Y_{i}(v_i)\neq Y_{i}(u_i)]$.  Let us remark, here, that there
is no standard way of computing $Pr[Y_{i}(v_i)\neq Y_{i}(u_i)]$. In the next section where
we study the application of the above schema on $G(n,d/n)$ we choose $t$ such that the
computation of the marginal $Pr[Y_{i}(v_i)\neq Y_{i}(u_i)]$ can be carried out efficiently
by using a {\em dynamic programming algorithm}.

As far as the accuracy is concerned we have the following results.

\begin{proposition}\label{proposition:SchemaAccRel}
For the counting schema it holds that
$$
\frac{1}{n}|\log {\cal Z}-\log Z(G,k)|\leq \frac{2}{n} 
\displaystyle \sum_{i=0}^{r-1}
\frac{|Pr[X_{i}(v_i)\neq X_{i}(u_i)]-Pr[Y_{i}(v_i)\neq Y_{i}(u_i)]|}
{Pr[X_{i}(v_i)\neq X_{i}(u_i)]},
$$
when each of the summands on the r.h.s. is sufficiently small.
\end{proposition}

\noindent
The proof of Proposition \ref{proposition:SchemaAccRel} appears in Section \ref{sec:thrm:SchemaAccRel}.

So as to show that the estimation $\log {\cal Z}$ is accurate, we work as follows:
We derive a constant lower bound for $Pr[X_{i}(v_i)\neq X_{i}(u_i)]$, which is used
to for the denominator in Proposition \ref{proposition:SchemaAccRel}. Then, we show that
$Pr[X_{i}(v_i)\neq X_{i} (u_i)]$ and $Pr[Y_{i}(v_i)\neq Y_{i}(u_i)]$ are asymptotically
equal. There, we use the following proposition.

\begin{proposition}\label{proposition:count-accuracy}
For $0 \leq i \leq r-1$ it holds that
\begin{displaymath}
\begin{array}{l}
\displaystyle |Pr[X_i(v_i)\neq X_{i}(u_i)]-Pr[Y_{i}(v_i)\neq Y_{i}(u_i)]| \leq 
\\ \vspace{-.3cm}\\
\qquad \qquad  \displaystyle  
\leq \sum_{j=0}^{r_i-1}C_{ij}
\max_{\sigma, \tau, \in \Omega(G_{ij}, k)} 
\left\{
||\mu_{i,j}(\cdot|\sigma_{v_{ij}})-\mu_{i,j}(\cdot|\tau_{v_{ij}}) ||_{\Psi_i\cup\{u_{ij}\}}
+
||\mu_{i,j}(\cdot|\sigma_{v_{ij}})-\mu_{i,j}(\cdot|\tau_{v_{ij}}) ||_{\{u_{ij}\}}
\right\},
\end{array}
\end{displaymath}
where $C_{ij}=\max_{s,t \in [k]}\left 
\{(Pr[X_{i,j}(u_{i,j})=s, X_{i,j}(v_{i,j})=t])^{-2} \right \}$
and $r_i$ is the number of terms in the sequence ${\cal G}(G_i)$.
\end{proposition}

\noindent
The proof of Proposition \ref{proposition:count-accuracy} is given in Section
\ref{sec:count-accuracy}.

\subsection{Remarks on the Spatial Conditions}\label{sec:RmrkSpatial}
It is interesting to discuss the implications of the spatial mixing conditions
required by Proposition \ref{proposition:SchemaAccRel} and Proposition \ref{proposition:count-accuracy}.
If every $C_{ij}$ in Proposition \ref{proposition:count-accuracy} is a sufficiently small
constant, which will be the case here, then the spatial mixing condition can be
summarized as follows: 
$$
\frac{1}{n}|\log {\cal Z}-\log Z(G,k)|\leq 
f(G,t) \cdot \max_{i,j,x, \sigma, \tau}
||\mu_{ij}(\cdot|\sigma_x)- \mu_{ij}(\cdot|\tau_x)||_{\Lambda}, \nonumber
$$
where $f(G,t)$ is a quantity that grows linearly with the number of terms in
both sequences ${\cal G}(G)$ and ${\cal G}(G_i)$  and $\Lambda\subset V$ is an
appropriate defined region in $G$. Then, a sufficient condition
for the counting schema to be accurate is that, for every
$0\leq i\leq r$ and $0 \leq j\leq r_i$ we have

\begin{equation}\label{eq:condition4accurancy}
\max_{x\in V}\max_{\sigma_x, \tau_x \in [k]^{\{x\}}} 
||\mu_{ij}(\cdot|\sigma_{x})- \mu_{ij}(\cdot|\tau_{x})||_{L(\{x\},t)}\leq
 \exp\left(-a\cdot t\right)
\end{equation}
for sufficiently large $a>0$.  Another expression for the condition in 
(\ref{eq:condition4accurancy}) can be derived by using the following 
(standard) lemma.

\begin{lemma}\label{lemma:reconstruction}
For any graph $G=(V,E)$ and $k$, let $\mu$ be the Gibbs distribution 
of its  $k$-colourings. For every $x\in V$ and $\Lambda \subseteq V$ it holds
\begin{displaymath}
\max_{\sigma_x, \tau_x \in [k]^{\{x\}}} 
||\mu(\cdot|\sigma_{x})- \mu(\cdot|\tau_{x})||_{\Lambda}
\leq 2k \cdot
\sum_{A\in [k]^{\Lambda}}
\mu(A)\cdot ||\mu(\cdot|A)-
\mu(\cdot)||_x.
\end{displaymath}
\end{lemma}

\noindent
For a proof Lemma \ref{lemma:reconstruction} see in Section \ref{sec:reconstruction}.

In the light of the above lemma  and for $k$ constant the condition in (\ref{eq:condition4accurancy}) 
is equivalent to  the following one: For  $0\leq i\leq r$ and $0\leq i\leq r_i$ 
\begin{equation}\label{eq:extrimality}
\max_{x\in V}\max_{\sigma_x, \tau_x \in [k]^{\{x\}}}  
\sum_{A\in [k]^{L(\{x\},t)}}
\mu_{ij}(A)\cdot 
||\mu_{ij}(\cdot|A)-\mu_{ij}(\cdot)||_{x}\leq \exp(-a'\cdot t),
\end{equation}
for appropriate $a'>0$. What the condition in (\ref{eq:extrimality}) implies is that
a ``typical'' colouring of $L(\{x\}, t)$ in $G_{ij}$ should have small impact on the Gibbs
 marginal on $x$.

\subsection{Bounds for Spatial Correlation decay}\label{sec:StmntSMBounds}
In this section, we provide the method that we use to derive an upper bound for
the quantities that express spatial correlation decay in Proposition \ref{proposition:count-accuracy},
i.e. $||\mu_{ij}(\cdot|\sigma_{x})- \mu_{ij}(\cdot|\tau_{x})||_{\Lambda}$, for
$x\in V$ and  $\Lambda \subset V$. The derivation of these bounds are of independent
interest from the discussion in the Section \ref{sec:RmrkSpatial}. The method is
based on  the well-known ``{\em disagreement percolation}'' coupling construction,
from  \cite{disagreement-percolation}.

Consider a configuration space on the vertices of $G$ such that each vertex $v\in V$ is
set either {\em disagreeing} or {\em non-disagreeing}. In such a configuration, we call
{\em path of disagreement} any simple path which has all its vertices disagreeing. Given
an integer $s$ and $w\in V$ we let  ${\cal P}_{s,w}$ be the {\em  product measure } under
which each vertex $v \in V\backslash\{w\}$ of degree $\Delta(v)<s$  is disagreeing with
probability  $\frac{1}{s-\Delta(v)}$ and non-disagreeing with the
remaining probability. If $s\leq \Delta(v)$, then $v$ is disagreeing with probability 1.
The vertex $w$ is set disagreeing with probability 1, regardless of its degree.
Using the above concepts we show the following result.

\begin{theorem}\label{thrm:UNStructPercBnd}
Consider the graph $G=(V,E)$, $v \in V$,  $\Lambda\subseteq V$ and an integer $k>0$. Let
$\mu$ denote the Gibbs distribution of the $k$-colourings of $G$. Also, let ${\cal P}_{k,v}$
denote the product measure  defined above. It holds that
$$
\max_{\sigma_v, \eta_v \in [k]^{\{v\}}}
||\mu(\cdot|\sigma_v) -\mu(\cdot|\eta_{v})||_{\Lambda}\leq 
{\cal P}_{s,v}[\textrm{
$\exists$ path of disagreement connecting $\{v\}$ and $\Lambda$}].
$$
\end{theorem}

\noindent
The proof of Theorem \ref{thrm:UNStructPercBnd} is given in Section \ref{sec:DisPercBounds}.

\begin{figure}
\centering
\begin{minipage}{0.24\textwidth}
	\centering
		\includegraphics[width=0.9\textwidth]{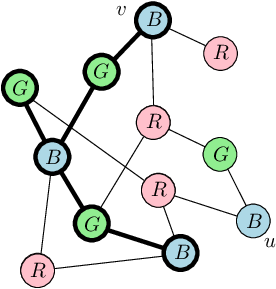}
		\caption{Colouring  $X$.}
	\label{fig:Gi0}
\end{minipage}
\begin{minipage}{0.24\textwidth}
	\centering
		\includegraphics[width=0.9\textwidth]{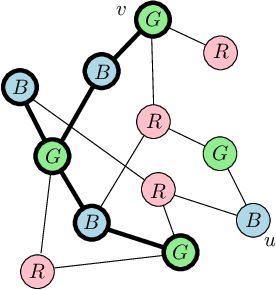}
		\caption{Colouring $Y$.}
	\label{fig:Gi-1}
\end{minipage}
\end{figure}

Roughly speaking, we bound $||\mu(\cdot|\sigma_{v})- \mu(\cdot|\eta_{v})||_{\Lambda}$, in
Theorem \ref{thrm:UNStructPercBnd}, by working as follows: We use coupling, i.e. we couple
$X,Y$ two random colourings of $G$ that assign the vertex $x$ colour $\sigma_v$ and $\eta_v$,
respectively. Then, by Coupling Lemma \cite{coupling-lemma} we have that 
$$
||\mu(\cdot|\sigma_{v})- \mu(\cdot|\eta_{v})||_{\Lambda}\leq Pr[X(\Lambda)\neq Y(\Lambda)].
$$
The coupling of $X,Y$ is done by specifying what $Y$ is, given $X$. In particular, given $X$,
we let $G_X$ denote the maximal {\em connected} subgraph of $G$ which contains the vertex $v$
and vertices from the colour classes specified by $\sigma_v$ and $\eta_v$ in the colouring $X$.
Then, we derive $Y$ as follows: For every vertex $u\notin G_X$ it holds that $Y(u)=X(u)$.
For $u\in G_X$ if $X(u)=\sigma_x$, then $Y(u)=\tau_x$ and the other way around\footnote{I.e.
if $X(u)=\tau_x$, then $Y(u)=\sigma_x$.}. In Figures \ref{fig:Gi0} and \ref{fig:Gi-1} we
illustrate this coupling, e.g. $\sigma_v=$``Blue'' and $\eta_v=$``Green''.

It is not hard to see that in the above coupling $X, Y$ disagree only on the colour
assignments for the vertices in $G_X$. That is 
$$Pr[X(\Lambda)\neq Y(\Lambda)]=Pr[\exists \Lambda'\subseteq\Lambda: \Lambda'\subseteq G_X \textrm{ in the coupling}].$$ 
Of course, bounding the probability term on the r.h.s. of the inequality above is
not a trivial task. However, we show that the above process (of getting $G_X$) is
stochastically dominated by an independent process, i.e. disagreement percolation.
That is, we show that
$$
Pr[\exists \Lambda'\subseteq\Lambda: \Lambda'\subseteq G_X \textrm{ in the coupling}]\leq {\cal P}_{s,v}[\textrm{ $\exists$ path of disagreement connecting $\{v\}$ and $\Lambda$}].
$$

\section{Application to $G(n,d/n)$}\label{sec:GnpAppMain}

In this section we show Theorem \ref{thrm:GnpApp}, Corollary \ref{theorem:free-nrg}
and Corollary \ref{thrm:verification}. For technical reasons, which we discuss 
later, we require the following sequence of subgraphs.\\ \vspace{-.3cm}

\noindent
{\bf Sequence of subgraphs ${\cal G}(G_{n,d/n})$:} Let $r$ be the greatest index in
${\cal G}(G_{n,d/n})$, e.g. ${\cal G}(G_{n,d/n})=G_0,\ldots, G_r$. The term $G_0$ is
an edgeless  graph. Let $R$ be the set of all edges in $G_{n,d/n}$ that do not belong
to a cycle of length smaller than \smallcycle  but they are incident to some vertex
that belongs to such a cycle. There is an index $i_0$ such that for every $i\geq i_0$,
$G_i$ differs from $G_{i-1}$ in some edge from $R$ while for $i<i_0$ no edge from the
set $R$ appears in $G_i$.

For $0\leq i \leq r$ consider that the sequence of subgraphs ${\cal G}(G_i)$ defined as
follows: $G_{i,0}$ is derived by $G_i$ by deleting all the edges that connect the sets
of vertices $L(\Psi_i, t)$ and $L(\Psi_i, t+1)$ where $t =\frac{\log n }{2\log d}$. 
\\ \vspace{-.2cm}

Typically we are in the case where $k$, the number of colours, is smaller than the maximum degree
of $G(n,d/n)$\footnote{The maximum degree in $G_{n,d/n}$  is
$\Theta\left(\frac{\log n}{\log\log n} \right )$ w.h.p. (see \cite{janson})}. Then,
there can be situations where  $(C_{i,j})^{-1}$ (defined in Proposition \ref{proposition:count-accuracy})
and $Pr[X_{i}(v_i)\neq X_{i}(u_i)]$ are very small. According to Proposition \ref{proposition:count-accuracy},
this can increase the error dramatically.  The analysis implies that these situations
arise when the vertices that are involved, i.e.  $v_i,u_i$, or $v_{ij}, u_{ij}$, have
large degrees and belong to small cycles at the same time.  It is easy to see that
choosing ${\cal G}(n,d/n)$ as we describe above, we avoid such undesirable situations
for any $i<i_0$. Furthermore, the terms $Pr[X_{i,0}(v_i)\neq X_{i,0}(u_i)]$ for
$i\geq i_0$ are too few, i.e. $O(n^{0.3})$, and it turns out that each of them is bounded away
from zero. This implies that their contribution to $\log(Z(G(n,d/n)))$ is negligible.

Setting the parameter $t=\frac{\log n }{2\log d}$, the component in $G_{i,0}$ which
contains $\{v_i,u_i\}$  is w.h.p. a tree with $O(\log n)$ extra edges, for every
$0\leq i<i_0$. This  allows the computation of every Gibbs marginal in polynomial-time.
To be more specific we work as follows:

\paragraph{Computing Probabilities}. The probability  term $Pr[Y_{i}(v_i) \neq Y_{i}(u_i)]$,
for $0\leq i<i_0$, can be computed by using Dynamic Programming (D.P.). More specifically,
using DP we can compute exactly the number of list colourings of a tree $T$. In the list
colouring  problem every vertex $v\in T$ has a set $List(v)$ of valid colours, where
$List(v)\subseteq [k]$ and $v$ only receives a colour in $List(v)$. For a tree on $l$
vertices, using dynamic programming we can compute exactly the number of list colourings
in time $lk$.

For $0\leq i<i_0$, the connected component in $G_{i,0}$ that contains $\{v_i, u_i\}$ is a
tree with at most $\Theta(\log n)$ extra edges w.h.p. For such component we can consider
all the $k^{O(\log n)}$ colourings of the endpoints of the extra edges and for each of
these colourings recurse on the remaining tree. Since in our case $k$ is  constant,
$k^{O(\log n)}=n^{O(1)}$. It follows that  the number of list colourings of the connected
component, in $G_{i,0}$,  that contains $\{v_i, u_i\}$ can be counted in polynomial time
for every $i$. This is sufficient for computing  $Pr[Y_{i}(v_i)\neq Y_{i}(u_i)]$ 
efficiently\footnote{A similar DP approach is also used in \cite{old-GnpSampling} and 
\cite{TCS-sampling}.}.

The pseudocode of the counting schema for the case of $G(n,d/n)$ follows. \\
\vspace{-.2cm}

\noindent
{\bf Counting Schema $G(n,d/n)$}\\ \vspace{-.65cm} \\
\rule{\textwidth}{1pt} \\
{\em Input}: $G(n,d/n)$, $k$\\ 
\hspace*{0.3cm} Compute the set of edges $R$.\\
\hspace*{0.3cm} If $|R|>n^{0.3}$, compute $\log(Z(G_{n,d/n},k))$ by {\em exhaustive enumeration}.\\
\hspace*{0.3cm} Compute the sequence of subgraphs ${\cal G}(G_{n,d/n})$.\\
\hspace*{0.3cm} Set ${\cal Z}=1$\\ 
\hspace*{0.3cm} For $0<i<r-|R|$ do
\begin{itemize}
\item Compute the exact value of $Pr[Y_{i}(v_i)\neq Y_{i}(u_i)]$.
\item Set ${\cal Z}={\cal Z}\cdot Pr[Y_{i}(v_i)\neq Y_{i}(u_i)]$.
\end{itemize}
\hspace*{0.3cm} End for.\\
\hspace*{0.3cm} Set ${\cal Z}={\cal Z}\cdot k^n$.\\ 
{\em Output}: $\displaystyle \log \left ({\cal Z}\right) /n$.\\
\rule{\textwidth}{1pt} \\ \vspace{-.3cm}\\

\noindent
Observe that, above,  implicitly we set $Pr[Y_{i}(v_i)\neq Y_{i}(u_i)]=1$ for $i\geq i_0$.
It turns out that the error introduced by working this way is negligible. Theorem \ref{thrm:GnpApp} 
follows as a corollary of the following two propositions.

\begin{proposition}\label{thrm:GnpAccuracy}
Let $\epsilon>0$ be a fixed number and let $d$ be sufficiently large. For $k\geq (2+\epsilon)d$ the
counting schema computes an  $n^{-b}$-approximation of $\log Z(G(n,d/n), k)$, with probability at
least $1-n^{-a}$, over the graph instances and $a,b>0$ depend on $k$.
\end{proposition}

\noindent
The proof of Proposition \ref{thrm:GnpAccuracy} appears in  Section \ref{sec:thrm:GnpAccuracy} 
and makes a heavy use of Theorem \ref{thrm:UNStructPercBnd}.

\begin{proposition}\label{thrm:GnpComplexity}
There are real constants $h,s>0$  such that the time complexity for the counting schema to compute 
$\log Z(G(n,d/n), k)$ is $O(n^s)$, with probability at least $1-n^{-h}$, over the graph instances.
\end{proposition}

\begin{myproof}
The theorem follows directly from the paragraph, ``Computing Probabilities'', above.
\end{myproof}

\subsection{Proof of Proposition \ref{thrm:GnpAccuracy}}\label{sec:thrm:GnpAccuracy}
First we present a series of results that will be useful for the proof of Proposition
\ref{thrm:GnpAccuracy}. In all our results that follow we assume that 
$\epsilon>0$ is  a fixed number and $d>0$ is sufficiently large, i.e. $d>d_0(\epsilon)$.

\begin{proposition}\label{proposition:spatial-gnp}
Consider the measure ${\cal P}_{k,x}$ w.r.t. $G(n,d/n)$, for $k\geq(2+\epsilon) d$ and some vertex
$x$ in the graph. For a set of vertices $\Psi$,  let $D^{(l)}$ denote the number of
paths of disagreement between $x$ and $\Psi$, of length at least $l$, for any integer
$l=O(\log n)$. Then, there exists a real $\gamma=\gamma(k)>1$ such that
\begin{equation}\label{eq:spatial-gnp}
Pr[D^{(l)}>0]\leq \frac{8}{\epsilon}\cdot \frac{|\Psi|}{n} \gamma^{-l},
\end{equation}
where $|\Psi|$ is the cardinality of $\Psi$. The probability term above, is w.r.t ${\cal P}_{k,x}$ and the graph instances.
\end{proposition}

\noindent
The proof of Proposition \ref{proposition:spatial-gnp} appears in Section \ref{sec:proposition:spatial-gnp}. 
Also, from the proof of Proposition \ref{proposition:spatial-gnp} it is direct to deduce
the following corollary.

\begin{corollary}\label{crlr:increasing-prop}
The bound for the probability in (\ref{eq:spatial-gnp}) holds even if we remove an arbitrary
set of edges of $G(n,d/n)$.
\end{corollary}

\noindent 
The following lemma is standard.  
We denote by $C_{l}$ the number of cycles of length at most $l$. Also, we 
remind the reader that the set $R$ is the set of 
edges of $G(n,d/n)$ that do not belong to a cycle of length smaller than 
\smallcycle  but they are incident to a vertex that belongs to such a cycle.

\begin{lemma}\label{lemma:REdges}
With probability at least $1-n^{-0.19}$, the following holds: (A) $|R| \leq n^{0.3}$. 
(B) $C_{l}\leq n^{0.3}$, for $l=$\smallcycle. (C) After removing  the edges in $R$ from $G_{n,d/n}$,
each of the cycles of length less than \smallcycle  becomes isolated from the rest of 
the graph.
\end{lemma}

\noindent
For completeness we present the proof of Lemma \ref{lemma:REdges} in  Section \ref{sec:lemma:REdges}.

\begin{lemma}\label{lemma:GnpCij}
For ${\cal G}(G(n,d/n))$,  ${\cal G}(G_i)$
as defined  in  Section \ref{sec:GnpAppMain} and  for constant $k\geq (2+\epsilon)d$, the following holds:
\begin{eqnarray}\label{eq:CijBounds}
Pr[C_{i,j}<2k^4,\: \textrm { for $0\leq i< i_0$, $0\leq j\leq r_i$}]\geq 1-n^{-\frac{\log \gamma}{11\log d}},
\end{eqnarray}
where $C_{ij}$, $\gamma$ are defined in the statements of Proposition \ref{proposition:count-accuracy} and Proposition
\ref{proposition:spatial-gnp}, respectively.
\end{lemma}

\begin{myproof}
Let $X_{i,j}$ be a random colouring of $G_{i,j}$. We remind the reader that 

$$C_{ij}=\max_{s,t \in [k]}\left \{(Pr[X_{i,j}(u_{i,j})=s, X_{i,j}(v_{i,j})=t])^{-2} \right \}.$$
We  show that $C_{i,j}$ is reasonably small by comparing $Pr[X_{i,j} (u_{i,j})=s|X_{i,j}(v_{i,j})=t]$
with $Pr[X_{i,j}(u_{i,j})=s]=1/k$ and by showing that these two probability terms do not differ much.
In particular,  we have 
\begin{eqnarray}\label{eq:1941a}
|Pr[X_{i,j}(u_{i,j})=s|X_{i,j}(v_{i,j})=t]-Pr[X_{i,j}(u_{i,j})=s]|\leq
\max_{\sigma,\eta \in [k]^{\{v_{i,j}\}}} 
||\mu_{ij}(\cdot|\sigma)- \mu_{ij}(\cdot|\eta)||_{u_{ij}}.
\end{eqnarray}
Then, we show that with probability at least $1-n^{-\frac{\log \gamma}{11\log d}}$  for $0\leq i<i_0$ 
and $0\leq j\leq r_i$ it holds that

\begin{eqnarray}\label{eq:1941b}
\max_{\sigma,\eta \in [k]^{\{v_{i,j}\}}} 
||\mu_{ij}(\cdot|\sigma)- \mu_{ij}(\cdot|\eta)||_{u_{ij}}
\leq \frac{1}{10k}.
\end{eqnarray}
Given the above, it is straightforward to verify (\ref{eq:CijBounds}) by using (\ref{eq:1941a})
and (\ref{eq:1941b}).
Then, the lemma follows.

We are going to use Theorem \ref{thrm:UNStructPercBnd} to prove (\ref{eq:1941b}). 
For a pair of adjacent vertices $x,y$ in the graph let $D_{x,y}$ denote the number
of paths of disagreement that start from $x$ and end in $y$ but they do not use
the edge $\{x,y\}$. Also, we let $\varrho_{x,y}={\cal P}_{k,x}[D_{x,y}>0]$. Finally,
given some  integer $s>1$ we  let $D^{(s)}_{x,y}$ denote the number of paths of
disagreement that start form $x$, end in $y$ and their length is at least $s$.
Similarly, let $\varrho^{(s)}_{x,w}={\cal P}_{k,x}[D^{(s)}_{x,y}>0]$.

\remove{
Theorem \ref{thrm:UNStructPercBnd} implies the following: 
\begin{eqnarray}\label{eq:1941c}
\max_{\sigma,\eta \in [k]^{\{v_{i,j}\}}} 
||\mu_{ij}(\cdot|\sigma)- \mu_{ij}(\cdot|\eta)||_{u_{ij}}
\leq {\cal P}_{k,v_{i,j}}[D_{v_{i,j},u_{i,j}}>0]=\varrho_{v_{i,j},u_{ij}}.
\end{eqnarray}

\noindent
Observe that $\varrho_{v_{i,j},u_{ij}}$ depends only on the structure of $G_{ij}$. That
is, given the graph $G_{i,j}$  its value is fully specified. We are going to derive an
appropriate  tail bound for $\varrho_{v_{i,j},u_{ij}}$. 
}

Let $e=\{x,y\}$ be a random edge in $G(n,d/n)$ conditional that the shorter cycle that contains it is of
length at least \smallcycle. Let $e'=\{x',y'\}$ be a randomly  chosen edge in $G(n,d/n)$. 
It holds that

\begin{eqnarray}\label{eq:1941d}
 E[\varrho_{x,y}]\leq \frac{1}{\psi}E[\varrho^{(l)}_{x',y'}],
\end{eqnarray}
where $l$ denotes the distance between the vertices $x$ and $y$. Also,  $\psi$ is the
probability that a randomly chosen edge in $G(n,d/n)$ does not belong to a cycle shorter
than \smallcycle.   It is straightforward to show that $\psi=1-o(1)$. Using Proposition
\ref{proposition:spatial-gnp} and 
the fact that $l\geq$\smallcycle we have that

\begin{eqnarray}\label{eq:1941e}
 E[\varrho^{(l)}_{x',y'}] \leq \frac{8}{\epsilon}n^{-\left(1+\frac{\log \gamma}{10\log d}\right)}.
\end{eqnarray}

\noindent
From (\ref{eq:1941d}) and (\ref{eq:1941e}) we get that $E[\varrho_{x,y}]\leq \frac{10}{\epsilon}n^{-\left(1+\frac{\log \gamma}{10\log d}\right)}$. 
From  Markov's inequality we get that
\begin{eqnarray} \nonumber
 Pr\left[\varrho_{x,y}\geq \frac{1}{10k}\right]\leq \frac{100k}{\epsilon}n^{-\left(1+\frac{\log \gamma}{10\log d}\right)}.
\end{eqnarray}

\noindent
Let ${L}$ be number of  edges $\{x,y\}$ in $G(n,d/n)$ such that the shortest cycle that 
contains each of them is of length at least \smallcycle $\:$   and $\varrho_{x,y}\geq \frac{1}{10k}$.
Using the linearity of expectation, it is straightforward to show that
$E[{L}]\leq \frac{60dk}{\epsilon}n^{-\frac{\log \gamma}{10\log d}}$.
Applying, Markov's inequality we get that
\begin{eqnarray}\label{eq:1942g}
 Pr[{L}>0]\leq \frac{60dk}{\epsilon}n^{-\frac{\log \gamma}{10\log d}}.
\end{eqnarray}

\noindent
Observe that the probability for path between two vertices to be a path of disagreement
is an increasing function of the degrees of its vertices (when $k$ is fixed). From this
observation and (\ref{eq:1942g}) we have that for every $v_{i,j}$ and $u_{i,j}$ it holds that
$\varrho_{v_{i,j},u_{ij}}\leq 1/(10k)$ with probability at least 
$1-\frac{60dk}{\epsilon}n^{-\frac{\log \gamma}{10\log d}}$. The lemma 
follows by  using Theorem \ref{thrm:UNStructPercBnd}, i.e. it holds that
\begin{eqnarray}\nonumber 
\max_{\sigma,\eta \in [k]^{\{v_{i,j}\}}} 
||\mu_{ij}(\cdot|\sigma)- \mu_{ij}(\cdot|\eta)||_{u_{ij}}
\leq {\cal P}_{k,v_{i,j}}[D_{v_{i,j},u_{i,j}}>0]=\varrho_{v_{i,j},u_{ij}}.
\end{eqnarray}
\end{myproof}

\begin{lemma}\label{lemma:GnpPrLarge}
Let $\gamma$ be as
in the statement of Proposition \ref{proposition:spatial-gnp}. For ${\cal G}(G_{n,d/n})$
as defined in Section \ref{sec:GnpAppMain} and for $k\geq (2+\epsilon)$ the following holds:
\begin{itemize}
 
\item Let $I$ be the set such that $i\in I$, iff the edge $\{v_i,u_i\}$ does not belong
 to any cycle of length less than \smallcycle. With probability at least $1-n^{-\frac{\log \gamma}{22\log d}}$
 over the instances $G(n,d/n)$ it holds that 
 \begin{eqnarray}\label{eq:GnpPrLargeMostEdges}
 \left|Pr[X_{i}(u_i)\neq X_i(v_i]-\left(1-\frac{1}{k}\right)\right|\leq n^{-\frac{\log \gamma}{21\log d}}, \quad \forall i\in I.
 \end{eqnarray}

 \item Let $I'$ be the set such that $i\in I'$, iff the edge $\{v_i,u_i\}$ belongs to cycle
 of length less than \smallcycle. With probability at least $1-n^{-0.19}$ over the instances
 $G(n,d/n)$ it holds that 
 $$
 Pr[X_{i}(u_i)\neq X_i(v_i]=\Theta(1).
 $$
\end{itemize}
\end{lemma}

\begin{myproof}
First we consider the edges $\{v_i,u_i\}$ such that $i\in I$. There, we use the following fact.
\begin{eqnarray}\label{eq:1942a}
\left|Pr[X_{i}(u_i)\neq X_i(v_i]-\left(1-\frac{1}{k}\right)\right|&\leq &
\max_{\sigma,\eta \in [k]^{\{v_i\}}} ||\mu_{i}(\cdot|\sigma)- \mu_{i}(\cdot|\eta)||_{u_i} 
\leq {\cal P}_{k,v_{i}}[D_{v_i,u_i}>0], \nonumber
\end{eqnarray}
where $D_{v_i,u_i}$ is the number of paths of disagreement in $G(n,d/n)$ that connect $v_i$ and $u_i$ 
but they do not use the edge $\{v_i,u_i\}$.

As in the proof of Lemma \ref{lemma:GnpCij}, for the vertices $x',y'$ we let 
$\varrho_{x',y'}={\cal P}_{k,x'}[D_{x',y'}>0]$. We work in the same manner as in
the proof of Lemma \ref{lemma:GnpCij} to get tail bounds for $\varrho_{x',y'}$, 
i.e. we get the following: For a random edge $\{x,y\}$ such that the shortest
cycle that contains it is of length at least \smallcycle, it holds that

\begin{eqnarray}
 Pr\left[\varrho_{x,y}\geq n^{-\frac{\log \gamma}{20\log d}}\right]\leq \frac{10}{\epsilon}n^{-\left(1+\frac{\log \gamma}{20\log d}\right)}.
\end{eqnarray}
Let ${L}$ be number of  edges in $G(n,d/n)$ such that the shortest cycle that
contains each of them is of length at least \smallcycle and $\varrho_{x,y}\geq n^{-\frac{\log \gamma}{20\log d}}$. 
Using the linearity of expectation it is straightforward to show that 
$ E[{L}]\leq \frac{6d}{\epsilon}n^{-\frac{\log \gamma}{20\log d}}$.
Applying, Markov's inequality we get that

\begin{eqnarray}
 Pr[{L}>0]\leq \frac{6d}{\epsilon}n^{-\frac{\log \gamma}{20\log d}}.
\end{eqnarray}
It is immediate that (\ref{eq:GnpPrLargeMostEdges}) holds.

In the latter case, we consider $v_i$ and $u_i$ which belong to small cycle, i.e.
of length at most \smallcycle. Such a pair of vertices appears in the schema
only when we have removed from $G_{n,d/n}$ all the edges in $R$. By Lemma 
\ref{lemma:REdges} we have that with probability at least $1-n^{-0.19}$ the
removal of the edges in $R$ disconnects every small cycle from the rest of
$G_{n,d/n}$. Thus, for the second case, where $v_i,u_i$ belong to a small,
isolated cycle, $Pr[X_{i}(u_i)\neq X_i(v_i]$ is trivially lower bounded by
some constant, since $k\gg2$. The lemma follows.
\end{myproof}

\noindent
Using Lemma \ref{lemma:Count2Sample} and the previous lemmas, in this section,
we get the following corollary.

\begin{corollary}\label{cor-logGnp}
For $k\geq (2+\epsilon)d$, the log-partition function of the $k$-colourings of 
$G_{n,d/n}$ is $\Theta(n)$, w.h.p. 
\end{corollary}

\noindent
We have all the lemmas we need to show Proposition \ref{thrm:GnpAccuracy}.\\

\begin{propositionproof}{\ref{thrm:GnpAccuracy}}
Let ${\cal D}$ be the event that ``
(a)  $r\leq \rho=\frac{dn}{2}(1+n^{-1/3})$, 
(b) $\max_i\{ r_i\} \leq 10 dn^{1/2}\log n$,  
(c) $|R|\leq n^{0.3}$, 
(d) $\min_{i} \{Pr[X_i(v_i)\neq X_{i}(u_i)]\}=\Theta(1)$, 
(e) $\max_{i,j}(C_{i,j})\leq 2k^4$''.

We remind the reader that we denote with $r$ the number of terms in ${\cal G}(G(n,d/n))$, 
$r_i$ the number of terms in ${\cal G}(G_i)$, for every $G_i\in {\cal G}(G(n,d/n))$.

\begin{myclaim}
It holds that $Pr[{\cal D}]\geq 1-n^{-\beta}$, for some fixed $\beta>0$.
\end{myclaim}

\begin{myproof}
From all the previous results in Section \ref{sec:thrm:GnpAccuracy}, it suffices to show that
$\max_i\{ r_i\} \leq 5 dn^{1/2}\log n$ with sufficiently large probability. 

Clearly, $r_i$ is equal to the number of edges between $L\left(\Psi_i, \frac{\log n}{2\log d}\right)$
and $L\left(\Psi_i, \frac{\log n}{2\log d}+1\right)$ in $G_i$.  The number of vertices at distance
$\frac{\log n}{2\log d}$ from $\Psi$ is dominated by a Galton-Watson tree of $\frac{\log n}{2\log d}$
levels, with a number of offspring per individual distributed as in ${\cal B}(n,d/n)$ and the initial
population being 2. With standard arguments (e.g. see Theorem  6 in \cite{mossel-ising-gnp}), it holds
that with probability at least $1-n^{-3}$, the number of vertices at level $\frac{\log n}{2\log d}$ is
at most $9n^{1/2} \log n$. Clearly $r_i$ is at most the sum of degrees of these vertices. In turn,
this sum is dominated by a sum of $9n^{1/2} \log n$ independent ${\cal B}(n,d/n)$. It is direct to
derive that $r_i=10d n^{1/2} \log n$ with probability at least $1-n^{-3}$, by using Chernoff bounds.
The claim follows.
\end{myproof}

\noindent
By Proposition \ref{proposition:SchemaAccRel} we have that  
\begin{eqnarray}\label{eq:LEVEL-A-Bound}
E \left [  \frac{1}{n}|\log {\cal Z}-\log Z(G(n,d/n))| | {\cal D}\right] \leq 
\frac{2}{n} \sum_{i=0}^{\rho}E\left[\frac{|Pr[X_{i}(v_i)\neq X_{i}(u_i)]-
Pr[X_{i,0}(v_i)\neq X_{i,0}(u_i)]|}{Pr[X_{i}(v_i)\neq X_{i}(u_i)]}|{\cal D}\right],
\end{eqnarray}
where the expectation is over the graph instances $G(n,d/n)$.
Using Proposition \ref{proposition:count-accuracy}, we have that

\begin{eqnarray}\label{eq:LEVEL-B-Bound}
E\left[|\frac{Pr[X_i(v_i)\neq X_{i}(u_i)]-Pr[Y_{i}(v_i)\neq Y_{i}(u_i)]}{Pr[X_i(v_i)\neq X_{i}(u_i)]}||{\cal D}\right] \leq
C\cdot E\left[\sum_{j=0}^{r_i-1}C_{i,j}\cdot Q_{ij}|{\cal D}\right],
\end{eqnarray}
where $C>0$ is a fixed number and
$$Q_{i,j}=\max_{\sigma, \tau \in [k]^{\{v_{ij}\}}}
\left\{
||\mu_{i,j}(\cdot|\sigma)-\mu_{ij}(\cdot|\tau)||_{\Psi_i\cup\{u_{i,j}\}}
+||\mu_{i,j}(\cdot|\sigma)-\mu_{ij}(\cdot|\tau)||_{u_{ij}}
\right \}.$$

\noindent
Clearly (\ref{eq:LEVEL-B-Bound}) holds since, conditioning on event ${\cal D}$,  we have a
constant lower bound on $Pr[X_i(v_i)\neq X_{i}(u_i)]$, for every $i$. Also,  the following
holds: For any $i\leq i_0$ we have that 

\begin{eqnarray}\label{eq:LEVEL-C-Bound}
E\left[\sum_{j=0}^{r_i-1}C_{i,j}\cdot Q_{ij}|{\cal D}\right]\leq 2k^4\sum_{j=0}^{5 dn^{1/2}\log n}E[Q_{i,j}|{\cal D}],
\end{eqnarray}

\noindent
since from conditioning on ${\cal D}$, it holds that $r_i \leq 10 dn^{1/2}\log n$ and $C_{ij}<2k^4$.
Also, we have the following, 

\begin{eqnarray}\label{eq:ExpcEQijBound}
E[Q_{ij}|{\cal D}]\leq \frac{E[Q_{ij}]}{Pr[{\cal D}]}\leq \frac{35}{\epsilon}n^{-\left(1+\frac{\log \gamma}{10\log(d)}\right)} \qquad \qquad \qquad \mbox{[as $Pr[{\cal D}]>3/4$]},
\end{eqnarray}

\noindent
where the bound for $E[Q_{i,j}]$ in the  last inequality follows by working exactly as in Lemma
\ref{lemma:GnpCij}. The quantity $\gamma$ is defined in Proposition \ref{proposition:spatial-gnp}.
We remind the reader than  for $i<i_0$ the distance between $v_{i,j}$ and $u_{i,j}$ is
at least \smallcycle. 

Plugging into (\ref{eq:LEVEL-B-Bound}) the inequalities in (\ref{eq:ExpcEQijBound}) and (\ref{eq:LEVEL-C-Bound}),
we get the following: For sufficiently large $n$ and for any $i\leq i_0$ we have that
\begin{eqnarray}\label{eq:ExpErrMajor}
E\left[|\frac{Pr[X_i(v_i)\neq X_{i}(u_i)]-Pr[Y_{i}(v_i)\neq Y_{i}(u_i)]}{Pr[X_i(v_i)\neq X_{i}(u_i)]}||{\cal D}\right] \leq n^{-\frac{1}{2}-\frac{\log \gamma}{10\log(d)}}.
\end{eqnarray}

\noindent
From the pseudocode of the schema for $G(n,d/n)$ we have that for $i\geq i_0$ the schema estimates
$Pr[X_i(v_i)\neq X_{i}(u_i)]$ by assuming that they are 1. Assuming that the event ${\cal D}$ holds,
then, it is not hard to show that 

\begin{eqnarray}\label{eq:ExpErrMinor}
\frac{|Pr[X_i(v_i)\neq X_{i}(u_i)]-1|}
{Pr[X_i(v_i)\neq X_{i}(u_i)]}=\Theta(1)\qquad \textrm{ for $i\geq i_0$.}
\end{eqnarray}

\noindent
Plugging (\ref{eq:ExpErrMajor}) and (\ref{eq:ExpErrMinor}) into (\ref{eq:LEVEL-A-Bound}) we get that

\begin{eqnarray}
E \left [  \frac{1}{n}|\log {\cal Z}-\log Z(G(n,d/n))| | {\cal D}\right] \leq 
2n^{-\left(1/2+\frac{\log \gamma}{11\log d}\right)}.\nonumber
\end{eqnarray}

\noindent
Using Markov's inequality we get that 
\begin{eqnarray}\nonumber
Pr \left [  \frac{1}{n}|\log {\cal Z}-\log Z(G(n,d/n),k)|\geq n^{-1/4} | {\cal D}\right] \leq 
2n^{-\left(1/4+\frac{\log \gamma}{11\log d}\right)}.
\end{eqnarray}
The proposition follows from the above inequality and the fact that $Pr[{\cal D}]\geq 1-n^{-\beta}$, for fixed $\beta>0$.
\end{propositionproof}

\subsection{Proof of Proposition \ref{proposition:spatial-gnp}}\label{sec:proposition:spatial-gnp}

\noindent
For the proof of Proposition \ref{proposition:spatial-gnp}, we need the following result.

\begin{lemma}\label{lemma:prob-disagreementpath}
Consider the graph $G(n,d/n)$ and let $\pi$ be a permutation of $l+1$ vertices of $G_{n,d/n}$, for $0\leq l \leq \Theta(\log^6 n)$. 
Consider, also, the product measure ${\cal P}_{k,x_1}$ w.r.t. the graph $G(n,d/n)$,
where $x_1=\pi(1)$ and $k\geq (2+\epsilon)d$. Setting $\Gamma=1$ if $\pi$ is a path of 
disagreement, otherwise $\Gamma=0$, it holds that

\begin{displaymath}
E[\Gamma] \leq \left ( \frac{d}{n}\right)^l \cdot
\left(  \left( \frac{1}{(1+\epsilon/2)d} +d^{-20}\right )^l+2n^{-\log^4 n} \right),
\end{displaymath}
where the expectation is taken w.r.t. both ${\cal P}_{k,x_1}$ and $G(n,d/n)$.
\end{lemma}

\begin{myproof}
Call $\pi$ the path that corresponds to the permutation $\pi$, e.g. $\pi=(x_1, \ldots x_{l+1})$.
Let $I_{\pi}$ be the event that there exists the path  $(x_1,\ldots,x_{l+1})$ in $G_{n,d/n}$. It holds that

\begin{displaymath}
E[\Gamma]=\left (\frac{d}{n}\right)^l 
\cdot E[\Gamma| I_{\pi}],
\end{displaymath}
Let $Q_{\pi}$ denote the event that the vertices in $\pi$ have degree less than $\log^6 n$. Using Chernoff bounds
it is easy to show that $Pr[Q_{\pi}|I_{\pi}]\geq 1-n^{-\log^4(n)}$. Also, it holds that
\begin{displaymath}
\begin{array}{lcl}
E[\Gamma| I_{\pi}] &=& E[\Gamma| I_{\pi},Q_{\pi}]Pr[Q_{\pi}|I_{\pi}]+ E[\Gamma| I_{\pi},\bar{Q}_{\pi}]Pr[\bar{Q}_{\pi}|I_{\pi}]
\\ \vspace{-.3cm}\\
 &\leq& E[\Gamma| I_{\pi},Q_{\pi}]+n^{-\log^4(n)}.
\end{array}
\end{displaymath}

\noindent
It suffices to show that for  $0\leq l \leq \Theta(\log^6 n) $ and sufficiently large $n$ it holds that 
\begin{equation}\label{eq:Gnpexppath}
E[\Gamma | I_{\pi}, Q_{\pi}]\leq
\left( \frac{1}{(1+\epsilon/2)d} +d^{-20}\right )^l.
\end{equation}

\noindent
We show (\ref{eq:Gnpexppath}) by using induction on $l$. Clearly for $l=0$ the inequality in
(\ref{eq:Gnpexppath}) is true. Assuming that (\ref{eq:Gnpexppath}) holds for $l=l_0$, we will
show that it holds for $l=l_0+1$, as well.

Let $D_i$, denote the event that the vertex $x_i$ is disagreeing. It suffices to show that
\begin{eqnarray}\label{eq:target-MarProbDis}
 Pr[D_{l_0+1}|\wedge_{j=1}^{l_0}D_j,I_{\pi}, Q_{\pi}] \leq \frac{1}{(1+\epsilon/2)d} +d^{-20}.
\end{eqnarray}

\noindent
Using the law of total probability, we have that
\begin{eqnarray}\label{eq:prob-DisMarg}
 Pr[D_{l_0+1}|\wedge_{j=1}^{l_0}D_j,I_{\pi}, Q_{\pi}] &\leq&
 Pr[D_{l_0+1}|\wedge_{j=1}^{l_0}D_j,I_{\pi}, Q_{\pi}, \Delta_{l_0+1}=0]+ \nonumber \\ 
 &&+Pr[\Delta_{l_0+1}>0|\wedge_{j=1}^{l_0}D_j,I_{\pi}, Q_{\pi}],
\end{eqnarray}
where ${\Delta}_{l_0+1}$ is the number of edges that are incident  to $x_{l_0+1}$ and some vertex
in $\{x_1, \ldots, x_{l_0-1}\}$.

Given that all vertices in $\{x_1, \ldots,x_{l_0}\}$ are disagreeing, let $\delta_{i}$ be the
number of vertices in $V \backslash\{x_1, \ldots, x_{l_0}\} $ that are adjacent to $x_i$, for
$1\leq i\leq l_0$. If $\delta_{i}=t$, then all the possible subsets of $V \backslash \{x_1, \ldots, x_{l_0}\}$
with cardinality $t$ are  equiprobably adjacent to $x_i$.  This implies that the probability
for $x_{l_0+1}$ to be  adjacent  to $x_i$ is $\frac{E[{\delta}_{i}]}{n-l_0}$.  By the linearity 
of expectation  we have

\begin{equation}\label{eq:expectedindegree}
E[{\Delta}_{l_0+1}|\wedge_{j=1}^{l_0}D_j,I_{\pi}, Q_{\pi}]\leq \frac{1}{n-l_0}\sum_{s=1}^{l_0}E[{\delta}_{s}|\wedge_{j=1}^{l_0}D_j,I_{\pi},Q_{\pi}]\leq n^{-0.97},
\end{equation}
the last inequality follows from the fact that $l_0\leq \Theta(\log^6n)$ and all the expectations in 
the sum  are upper bounded by $\log^6 n$, due to conditioning on $Q_{\pi}$. By (\ref{eq:expectedindegree}) and Markov's 
inequality, we get that 

\begin{eqnarray}\label{eq:ProbIndegree}
Pr[\Delta_{l_0+1}>0|\wedge_{j=1}^{l_0}D_j,I_{\pi}, Q_{\pi} ] \leq n^{-0.97}.
\end{eqnarray}
Also, we have that
\begin{eqnarray}
\varrho&=& Pr[D_{l_0+1}|\wedge_{j=1}^{l_0}D_j,I_{\pi}, Q_{\pi}, \Delta_{l_0+1}=0]\nonumber\\
&\leq&
 \sum_{j=0}^{k-3}\frac{1}{k-2-j}{n \choose j}(d/n)^j(1-d/n)^{n-j}+ \sum_{j=k-2}^{n-2}{n \choose j}(d/n)^j(1-d/n)^{n-j} \nonumber \\
 &\leq& \frac{1}{(2+\epsilon)d/2} \sum_{j=0}^{(2+\epsilon)d/2}{n \choose j}(d/n)^j(1-d/n)^{n-j}+ \sum_{j=(2+\epsilon)d/2+1}^{n-2}{n \choose j}(d/n)^j(1-d/n)^{n-j} \nonumber\\
 &\leq& \frac{1}{(2+\epsilon)d/2} +\exp\left(-cd\right), \label{eq:varrhoBound}
 \end{eqnarray}
where $c=\log c'-1+1/c'$ and $c'=(1+\epsilon/2)$. The last inequality follows from Chernoff
bounds, i.e. Corollary 2.4 in \cite{janson}. Plugging (\ref{eq:varrhoBound}) and (\ref{eq:ProbIndegree})
into (\ref{eq:prob-DisMarg}), for large $d$ we get that
\begin{eqnarray}\nonumber
 Pr[D_{l_0+1}|\wedge_{j=1}^{l_0}D_j,I_{\pi}, Q_{\pi}]\leq  \frac{1}{(1+\epsilon/2)d} +d^{-20}.
\end{eqnarray}
That is, (\ref{eq:target-MarProbDis}) is true. The lemma follows.
\end{myproof}

\begin{propositionproof}{\ref{proposition:spatial-gnp}}
Consider an enumeration of all the permutations of $t\geq l$ vertices in $G(n,d/n)$ with first
the vertex $x$ and last some vertex of $\Psi$. Let $\pi_0(t),\pi_1(t),\ldots$ be the permutations
in the order they appear in the enumeration. Also, w.r.t. the graph $G(n,d/n)$, consider the
product measure  ${\cal P}_{k,x}$  as it is defined in the statement of  Theorem \ref{thrm:UNStructPercBnd}.
Let $\Gamma_i(t)$ be the random  variable such that

\begin{displaymath}
\Gamma_i(t)=\left\{
\begin{array}{lcl}
1 &\qquad & \textrm{the path that corresponds to $\pi_i(t)$ is a path of disagreement}\\
0 && \textrm{otherwise}.
\end{array}
\right.
\end{displaymath}

\noindent
Let, also, $\Gamma(t)=\sum_i \Gamma_i (t)$. 

Let ${\cal E}=1$  if the event {\em ``there is no path of disagreement that starts from $x$ and has
length larger than $t_0=\frac{10\log n}{\log (1.04)}$''} occurs and ${\cal E}=0$ otherwise. 
It holds that 

\begin{eqnarray}
{\cal P}_{k,x_1}\left [\sum_{t\geq l}\Gamma(t)>0 \right] 
&\leq& {\cal P}_{k,x_1}\left [\sum_{t\geq l}\Gamma(t)>0 
| {\cal E}=1 \right] {\cal P}_{k,x_1}[{\cal E}=1] +{\cal P}_{k,x_1}[{\cal E}=0] 
\nonumber \\ 
&\leq & {\cal P}_{k,x_1}\left [\sum_{l\leq t<t_0}\Gamma(l)>0
\right]+Pr[{\cal E}=0].\label{eq:1978A}
\end{eqnarray}

\noindent
For convenience, we let $\varrho={\cal P}_{k,x_1}\left [\sum_{t\geq l}\Gamma(t)>0 \right] $, 
$\varrho_1= {\cal P}_{k,x_1}\left [\sum_{l\leq t<t_0}\Gamma(l)>0\right]$ and $\varrho_2=Pr[{\cal E}=0]$.
The proposition follows by deriving an appropriate upper bound for $E[\varrho]$, where the expectation
is taken w.r.t. graph instances.  For this we bound appropriately $E[\varrho_1]$ and $E[\varrho_2]$
and use the following inequality (which follows from (\ref{eq:1978A}))
\begin{eqnarray}\label{eq:ExpcVarrhos}
 E[\varrho]\leq E[\varrho_1] + E[\varrho_2].
\end{eqnarray}

\noindent
It holds that 
\begin{eqnarray}
E[\varrho_1] &\leq & \sum_{l\leq t<t_0} \sum_iE[\Gamma_{i}(t)] \nonumber \\
&\leq & \sum_{l\leq t<t_0} \frac{|\Psi|}{n} d^t \cdot
\left(  \left( \frac{1}{(1+\epsilon/2)d} +d^{-20}\right )^t+2n^{-\log^4 n} \right), \nonumber
\end{eqnarray}
where in the last inequality we use  Lemma \ref{lemma:prob-disagreementpath} and  the fact
that between $x_1$ and $\Psi$ there are at most $|\Psi|\cdot n^{t-1}$ paths of length exactly
$t$. Since $t\leq \log^2 n$,  it is direct that 
\begin{eqnarray}
 E[\varrho_1]&\leq &\sum_{l\leq t<t_0} \frac{|\Psi|}{n} (1+\epsilon/4)^{-t} \leq \frac{4+\epsilon}{\epsilon}\frac{|\Psi|}{n} (1+\epsilon/4)^{-l}.  \label{eq:Varrho_1Bound}
\end{eqnarray}

\noindent
Observe that  ${\cal P}_{k,x_1}[{\cal E}=0]\leq {\cal P}_{k,x_1}\left [H(t_0)>0 \right]$,
where $H(t_0)$ denotes the number of paths of disagreement of length $t_0$ that start from vertex $x_1$.
Note that the paths that $H(t_0)$ counts do not necessarily end in $\Psi$. By Markov's inequality, 
we have that
\begin{eqnarray}
 {\cal P}_{k,x_1}[{\cal E}=0]&\leq & E_{\cal P}[H(t_0)].\nonumber
\end{eqnarray}
Clearly, the above implies that $E[\varrho_2] \leq E[H(t_0)] $, where the expectations is taken
w.r.t. both ${\cal P}_{k,x_1}$ and the graph instances. We use  Lemma \ref{lemma:prob-disagreementpath}
to bound $E[H(t_0)] $ and we get that
\begin{eqnarray}
 E[\varrho_2]&\leq & n^{t_0} \left(\frac{d}{n} \right)^{t_0}\left( \left(\frac{1}{(1+\epsilon/2)d}+d^{-20} \right)^{t_0}+2n^{-\log^4n}\right) \qquad \mbox{[from Lemma \ref{lemma:prob-disagreementpath}]} \nonumber \\
 &\leq & \left(\frac{1}{1+\epsilon/4}\right)^{\log^2 n}+n^{-\frac12\log^4n}. \label{eq:Varrho_2Bound}
\end{eqnarray}

\noindent
The proposition follows by plugging (\ref{eq:Varrho_1Bound}) and (\ref{eq:Varrho_2Bound}) into
(\ref{eq:ExpcVarrhos}).
\end{propositionproof}

\subsection{Proof of Corollary \ref{theorem:free-nrg}}\label{sec:thrm:free-nrg}

For proving the corollary we are going to use Lemma \ref{lemma:Count2Sample}. In particular, 
it suffices to have the following: W.h.p over $G(n,d/n)$ all but a vanishing fraction of the
probability terms $Pr[X(v_i)\neq X(u_i)]$ are within distance $o(1)$ from $\left (1-\frac{1}{k} \right)$.
Also, the remaining probability terms, i.e. those which are not close to $\left (1-\frac{1}{k} \right)$
are bounded well away from zero. 

The corollary follows immediately from Lemmas \ref{lemma:REdges}, \ref{lemma:GnpPrLarge}. 
That is, consider the sequence of subgraph ${\cal G}(G(n,d/n))$ we have for the counting algorithm. 
From Lemma \ref{lemma:GnpPrLarge} and Lemma \ref{lemma:REdges} we have that w.h.p. the situation 
is as follows: There is a set of indices $I$ such that for every $i\in I$ it holds that

\begin{eqnarray}\label{eq:FreeNRGGoodTerms}
 |Pr[X(v_i)\neq X(u_i)]-\left(1-\frac{1}{k}\right)|\leq n^{-\frac{\log \gamma}{21\log d}}.
\end{eqnarray}
For the rest indices, i.e. $i\notin I$ it holds that 
\begin{eqnarray}\label{eq:FreeNRGBadTerms}
 |Pr[X(v_i)\neq X(u_i)]-\left(1-\frac{1}{k}\right)|=\Theta(1).
\end{eqnarray}
From Lemma \ref{lemma:Count2Sample} we can write $\frac{1}{n}\log(Z(G(n,d/n),k))$ as follows:
\begin{eqnarray}
 \frac{1}{n}\log Z(G(n,d/n),k)&=&k+\frac{1}{n}\sum_{i=1}^{r}\log Pr[X(v_i)\neq X(u_i)] \nonumber\\
 &=&k+\frac{1}{n}\sum_{i\in I}\log Pr[X(v_i)\neq X(u_i)]+\frac{1}{n}\sum_{i\notin I}\log Pr[X(v_i)\neq X(u_i)], \nonumber
\end{eqnarray}
while from Lemma \ref{lemma:REdges} we get that w.h.p. $|I|\geq n-O(n^{3/10}\log n)$. 
%
We derive upper and lower bounds for $\frac{1}{n}\log Z(G(n,d/n),k)$ by working as follows:
\begin{eqnarray}
 \frac{1}{n}\log Z(G(n,d/n),k) &\leq& k+ \frac{|I|}{n}\left(\left(1-\frac{1}{k}\right)+ n^{-\frac{\log \gamma}{21\log d}} \right) +\frac{n-|I|}{n} \nonumber \\
 &\leq& k+ \frac{d}{2}\left(1-\frac{1}{k}\right) +2n^{-\frac{\log \gamma}{21\log d}},
\end{eqnarray}
where in the last inequality we used the lower bound for the cardinality of the set $I$. Working in exactly
the same manner we get the lower bound for $\frac{1}{n}\log Z(G(n,d/n),k)$. The corollary follows.

\subsection{Proof of Corollary \ref{thrm:verification}} \label{sec:thrm:verification}

\noindent
Consider the following sequence of subgraphs ${\cal G}(G_{n,d/n})$ (different than what we used previously):
The term-graph $G_0$ is edgless. There is an index $i_1$ such that for $0<i\leq i_1$, $G_i$ contains 
all the edges that belong to cycles of length at most \smallcycle  in $G_{n,d/n}$ and only these edges. 
We  refer to the cycle of length less than \smallcycle $\:$ as ``small cycles''.

Let $S(n,d)$ be the set of  instances of $G_{n,d/n}$ which have
(A) $\Theta(n)$ edges, (B) $i_1\leq \Theta(n^{0.3}\log n)$ and (C) each 
$B(v_i,\frac{\log n}{4\log(e^2d/2)})$ is either a tree or unicyclic. 

We are going to show that for every $G\in S(n,d)$ and every term $G_i\in {\cal G}(G)$ such
that $i\geq i_1$, we can verify in  polynomial time that

\begin{equation}\label{eq:VerifyStep1}
||\mu(\cdot|\sigma_{v_i})-\mu(\cdot|\eta_{v_i})||_{u_i} \leq n^{-\epsilon_1},
\end{equation}
where $\epsilon_1>0$ . Then the corollary follows by using standard arguments, i.e. 
from  Lemma \ref{lemma:Count2Sample} and from the fact that
$\left|Pr[X_i(u_i)\neq X_i(v_i)]-\left(1-\frac{1}{k}\right)\right|\leq  \max_{\sigma, \eta \in [k]^{\{v_i\}}}||\mu_i(\cdot|\sigma)-\mu_i(\cdot|\eta)||_{u_i}$.

The value of $\epsilon_1$ in (\ref{eq:VerifyStep1}) depends  on the function $h(n,k,d)$
and $i_1$. For $i<i_1$ it direct to see that $G_{i}$ is so simple that we can compute
$Pr[X_{u_i}\neq X_{v_{i}}]$ exactly. Theorem \ref{thrm:UNStructPercBnd} and Corollary
\ref{crlr:increasing-prop} suggest that

\begin{equation}\label{eq:VerifyStep2}
||\mu(\cdot|\sigma_{v_i})-\mu(\cdot|\eta_{v_i})||_{u_i} \leq 
{\cal P}_{k,v_i}[\textrm{
$\exists$ path of disagreement connecting $\{v_i\}$ and $\{u_i\}$}].
\end{equation}

\noindent 
where ${\cal P}_{k,v_i}$  is the product measure defined in Section \ref{sec:StmntSMBounds}
and it is taken w.r.t graph $G_{n,d/n}\backslash \{v_i, u_i\}$. For $i>i_1$ it holds 
that $dist(v_i,u_i)\geq \frac{\log n}{10\log(d)}$ in $G_{n,d/n}\backslash\{v_i,u_i\}$.
Consider, now,  the event   $$E_{v_i,c}=\textrm{``$\exists$ a path of disagreement that connects $v_i$ with 
$L(v_i, c\log n)$ in $G_{n,d/n}\backslash\{v_i,u_i\}$''}.$$

\noindent
For each pair $v_i$ $u_i$ define  
$$a_i=\min\left\{\frac{dist(v_i,u_i)}{\log n}, (4\log(e^2d/2))^{-1} \right\}.$$
Noting that, for fixed $c_1> c_2$ it holds that ${\cal P}_{k,v_i} [E_{v_i,c_1}] \leq {\cal P}_{k,v_i}[E_{v_i,c_2}]$, we get that

\begin{equation}\label{eq:VerifyStep3}
{\cal P}_{k,v_i}[\textrm{$\exists$ path of disagreement connecting 
$\{v_i\}$ and $\{u_i\}$ in $G_{n,d/n}\backslash\{v_i,u_i\}$}] \leq {\cal P}_{k,v_i}[E_{v_i,a_i}].
\end{equation}

\noindent
By (\ref{eq:VerifyStep1}) (\ref{eq:VerifyStep2}) and (\ref{eq:VerifyStep3}), we can verify (\ref{eq:VerifyStep1}) 
by using the criterion  ${\cal P}_{k,v_i}(E_{v_i,a_i})\leq n^{-\epsilon_1}$.  It remains to show 
that ${\cal P}_{k,v_i}(E_{v_i,a_i})\leq n^{-\epsilon_1}$, for $i\geq i_1$, can be verified in 
polynomial time. Let $T_{v_i, a_i}$ be the set of all simple paths that connect  $v_i$ to $L(v_i, a_i\log n)$, it holds that 
\begin{eqnarray}\label{eq:Prob2Compute}
{\cal P}_{k,v_i}[E_{v_i,a_i}]\leq \sum_{m\in T_{v_i, a_i}}{\cal P}_{k,v_i}
[\textrm{``$m$ is a path of disagreement''}].
\end{eqnarray}

\noindent
The computation of each probability term on the r.h.s. of the above inequality can be carried out in polynomial 
time. It suffices to show that w.h.p. the number of these terms is polynomially large.

Using Lemma 2.1 from \cite{TCS-sampling} we get that for
every $i>i_1$ the subgraph $B(v_i, a_i\log n)$ of $G_{n,d/n} \backslash\{v_i,u_i\}$, is a tree with at
most an extra edge, with probability at least $1-n^{-0.1}$. In this case, the number of simple paths
between $v_i$ and  $L(v_i, a_i\log n)$ is at most  $2|L(v_i, a_i\log n)|$. Also, with standard arguments
(e.g. see Theorem  6 in \cite{mossel-ising-gnp}), it holds that with probability at least $1-o(n^{-2})$,
$|L(v_i, a_i\log n) |\leq n^{0.26} \log n$, for every $i>i_1$.  That is, for every $i>i_1$, $|T_{v_i,a_i}|$ 
is polynomially  large with probability at least $1-2n^{-0.1}$. Thus, the probability term on the l.h.s.
of (\ref{eq:Prob2Compute}) can be computed efficiently, for any $i>i_1$, w.h.p.  

Using the arguments in the paragraph above and Lemma \ref{lemma:REdges} it is direct to show that
$Pr[G(n,d/n)\in S(n,d)]\geq 1-3n^{-0.1}$. Also, it is direct that we can decide whether  
$G(n,d/n)\in S(n,d)$ or not, efficiently. The corollary follows

\section{Bounds for spatial correlation decay - Proof of Theorem \ref{thrm:UNStructPercBnd}}\label{sec:DisPercBounds}

For some finite graph $G=(V,E)$ and some sufficiently large integer $k$, let  $\mu(\cdot)$ 
be the Gibbs distribution of the $k$-colourings of $G$. For $x\in V$, $\Lambda \subseteq V$
and $\sigma_x, \eta_x\in [k]^{\{x\}}$, we are interested in deriving upper bounds for 
following quantity 
\begin{equation}
||\mu(\cdot|\sigma_x) - \mu(\cdot|\eta_x)||_{\Lambda}. \nonumber
\end{equation}
Towards bounding the above quantity we introduce two random variables $X^{\sigma}$, $X^{\eta}\in [k]^V$
distributed as in $\mu(\cdot|\sigma_x)$ and $\mu(\cdot|\eta_{x})$, respectively. 
We couple $X^{\sigma}$ and $X^{\eta}$ and we use the following inequality from
the Coupling Lemma (see \cite{coupling-lemma}),
\begin{displaymath}
||\mu(\cdot|\sigma_x) - \mu(\cdot|\eta_x)||_{\Lambda}\leq
Pr[X^{\sigma}(\Lambda)\neq X^{\eta}(\Lambda) \textrm{ in the coupling}].
\end{displaymath}

\noindent
We provide a upper bound for the probability of the event  ``$X^{\sigma}(\Lambda)\neq X^{\eta}(\Lambda)$'' 
in the coupling, in terms of $k$ and the degrees of the vertices in $G$ by using ``{\em disagreement percolation}'',
\cite{disagreement-percolation}.
In Section \ref{sec:CouplingDefs} we describe the coupling between $X^{\sigma}$
and $X^{\tau}$.

\subsection{The coupling for the comparison}\label{sec:CouplingDefs}

Let $\Omega_{\sigma}$ and $\Omega_{\eta}$ denote the $k$-colourings of $G$ that assign the vertex
$x$ colour $\sigma_x$ and $\eta_x$, respectively.  For the coupling of $X^{\sigma}$ and $X^{\eta}$
we need to develop, first,  a bijection $T:\Omega_{\sigma}\to \Omega_{\eta}$ as follows:

Given $\xi\in \Omega_{\sigma}$, we let $G_{\xi}=(V_{\xi}, E_{\xi})$, {\em induced subgraph} of $G$,
be defined  as follows: In the colouring $\xi$, let $V_{\sigma}$ and $V_{\eta}$ be the colour classes
specified by the colours $\sigma_x$ and $\eta_x$, respectively. Then $G_{\xi}=(V_{\xi}, E_{\xi})$ is
the {\em maximal}, {\em connected} graph such that $x\in V_{\xi}$ and 
$V_{\xi}\subseteq V_{\sigma} \cup V_{\eta}$. That is, $G_{\xi}$ is the maximal, connected, induced 
subgraph of $G$ which contains $x$ and vertices only from the colour classes $V_{\sigma}$ and 
$V_{\eta}$, in the colouring $\xi$. Then, given $G_{\xi}$, we derive $T\xi$ by working as follows: 
For every vertex $u\notin G_{\xi}$ it holds that $\xi(u)=(T\xi)(u)$. For $u\in G_{\xi}$ if $\xi(u)=\sigma_x$, 
then $(T\xi)(u)=\eta_x$. Also, if  $\xi(u)=\eta_x$, then $(T\xi)(u)=\sigma_x$.

In Figures \ref{fig:Gi0} and \ref{fig:Gi-1}, in Section \ref{sec:StmntSMBounds}, we illustrate how does the
mapping $T$ work.  Of course, it is not direct that $T$ is a bijection. For this we provide the following lemma.

\begin{lemma}\label{lemma:bijection-T}
It holds that $T: \Omega_{\sigma}\to \Omega_{\eta}$ is a bijection. 
\end{lemma}
\begin{myproof}
For the colouring $\xi \in \Omega_{\sigma}$, consider $G_{\xi}=(V_{\xi},E_{\xi})$ as defined
above. We need to focus on three properties that $G_{\xi}$ has. First,  it is easy to see that 
$G_{\xi}$ should be bipartite (in the extreme case where $V_{\xi}=\{x\}$ we consider $G_{\xi}$
bipartite too). Second, $G_{\xi}$ is connected due to the way we consider it. Third,  the fact 
that  $G_{\xi}$ is  maximal implies the following: if 
$\partial V_{\xi}=\{ v\in V \backslash V_{\xi} | \{v, u \}\in E \;\textrm{for}\; u \in
V_{\xi} \}$, then  $\forall v\in \partial V_{\xi}$  it holds  $\xi_{v}\notin \{\sigma_{x}, \eta_{u }\}$.

Clearly $\xi$ specifies a proper $2$-colouring for the vertices of $G_{\xi}$ that uses only 
the colours $\sigma_{x}$ and $\eta_{x}$. In particular, let $p_1,p_2 \subseteq V_{\xi}$ be
the two parts of  $G_{\xi}$ and  w.l.o.g. assume that $x$ belongs to $p_1$. Then,  $\xi$ 
assigns to all the vertices in $p_1$ the colour $\sigma_{x}$ and to all the vertices in $p_2$
the colour $\eta_{x}$. In that terms, the mapping $T$ works as follows: For every vertex
$v \in V\backslash V_{\xi}$ to hold $(T\xi)_v=\xi_v$. For the remaining  vertices, i.e. those
that belong to $G_{\xi}$, the mapping $T$ swaps the colour assignments of the two parts of
$G_{\xi}$. 
%
First we show that $T$ maps every colouring of $\Omega_{\sigma}$ to $\Omega_{\sigma}$.

\begin{myclaim}\label{claim:proper-colouring}
For every $\xi \in \Omega_{\sigma}$ it holds that $(T\xi) \in \Omega_{\eta}$.   
\end{myclaim}
\begin{myproof}
It is direct that $(T\xi)_{x}=\eta_{x}$.  It remains to  show that $T\xi$ is a proper colouring  of $G $.

If $T\xi$ is a non proper colouring, then there should be , at least, two adjacent vertices (somewhere 
in $G $) having the same colour assignment. The swap of colour assignments that take place, when we apply
$T$ on $\xi$, involves only vertices in $V_{\xi}$. Thus if $(T\xi)$ is a non proper colouring, then the 
monochromatic pair of adjacent vertices has either both vertices in $V_{\xi}$ or one vertex in $V_{\xi}$
and the other in $\partial V_{\xi}$.

It is direct that swapping the colour assignments of the two parts of $G_{\xi}$, as these are specified 
by $\xi$, leads to a proper colouring  of $G_{\xi}$. Thus, in $T\xi$ there is no monochromatic pair whose
both vertices belong to $G_{\xi}$. Also, this swap of colourings cannot lead some vertex in $V_{\xi}$
to have the same colour assignment with some vertex in $\partial V_{\xi}$. This is due to the maximality
of $G_{\xi}$, i.e. the colouring $\xi$ cannot not  specify colour assignment that uses the colours
$\sigma_{x}$ and $\eta_{x}$ for  any vertex in $\partial  V_{\xi}$. Thus, for every 
$\xi\in \Omega_{\sigma}$, it holds that  $T\xi$ is a proper colouring of $G$. The claim follows.
\end{myproof}

\noindent
It remains to show that $T$ is a bijection. The next claim shows that 
$T$ is a surjective.

\begin{myclaim}\label{claim:range-mapping}
$T$ is surjective.
\end{myclaim}
\begin{myproof}
Let $\xi'$ be any member of $\Omega_{\eta}$. We are going to show that there exists $\xi \in \Omega_{\sigma}$ 
such that $T\xi=\xi'$. 

For the colouring $\xi'$, let $G_{\xi'}=(V_{\xi'},E_{\xi'})$ be  the maximal, connected bipartite subgraph of 
$G $ such that  $x \in V_{\xi'}$ and $\forall v \in V_{\xi'}$ it holds $\xi'_v \in \{\sigma_{x}, \eta_{x}\}$,
(i.e. $G_{\xi'}$ is derived in a similar way as $G_{\xi}$, above).

The colouring $\xi'$ specifies a proper $2$-colouring for $G_{\xi'}$  that uses only the colours $\sigma_{x}$
and $\eta_{x}$. Let $p_1, p_2 \subseteq V_{\xi}$ be the two parts of $G_{\xi'}$ and w.l.o.g. assume that
$\xi'$ assigns to all the vertices in $p_1$ the colour $\eta_{x}$ and to all the vertices in $p_2$ the 
colour $\sigma_{x}$. 

Consider the colouring $\xi$ which is derived by $\xi'$ by swapping the colour assignments of the two parts
of $G_{\xi'}$ while $\xi_v =\xi'_v$ for $v \in V \backslash V_{\xi'}$. With arguments similar to those in
the proof of Claim \ref{claim:proper-colouring} we can see that $\xi \in \Omega_{\sigma}$.  The claim 
follows by noting, additionally, that $T\xi=\xi'$.
\end{myproof}

\noindent
In the following claim we show that $T$ is one-to-one.

\begin{myclaim}\label{claim:mapping-1to1}
$T$ is one-to-one.
\end{myclaim}
\begin{myproof}
Assume that there are two colourings $ \xi^1, \xi^2 \in \Omega_{\sigma}$ such that $T\xi^1=T\xi^2=\xi^3$. We are 
going to show that it should hold $\xi^1=\xi^2$. 
For this, assume the opposite, i.e. $\xi^1 \neq \xi^2$. We consider the graphs $G_{\xi^1}$ $G_{\xi^2}$ and $G_{\xi^3}$, as in the proofs
of the two previous claims.  By the proofs of these claims we know that the graphs $G_{\xi^1}$, $G_{\xi^2}$ and $G_{\xi^3}$ 
have the same subset of vertices of $G$.

Thus, we conclude that the colourings $\xi^1$ and $\xi^2$ should differ only on the colour assignment of the vertices in the graph 
$G_{\xi^1}$. We remind the reader that this graph is a connected bipartite graph with $\xi^1$ and $\xi^2$ specifying proper 
2-colourings for $G_{\xi^1}$ which both using the colours $\{\sigma_{x}, 
\eta_{x}\}$. 

By assumption, the 2-colouring for $G_{\xi^1}$ that $\xi^1$ specifies is different than that of $T\xi^1$. The same holds for colouring of 
$\xi^2$ and $T\xi^2$. Since $T\xi^1=T\xi^2$ we deduce that there exist three different 2-colourings for $G_{\xi^1}$. 
There is a contradiction, here, since there can exist only two  2-colourings for $G_{\xi^1}$. The claim follows.
\end{myproof}

\noindent
Since the mapping $T:\Omega_{\sigma}\to \Omega_{\eta}$ is surjective (Claim \ref{claim:range-mapping})
and one-to-one (Claim \ref{claim:mapping-1to1}), it is a bijection.  The lemma follows.
\end{myproof}

\begin{lemma}\label{lemma:coupling}
There exists a coupling of $X^{\sigma}$ with $X^{\eta}$ such that
\begin{displaymath}
X^{\eta}=TX^{\sigma}.
\end{displaymath}
\end{lemma}
\begin{myproof}
The existence of the bijection $T$ implies that $|\Omega_{\sigma}|=|\Omega_{\eta}|$. Thus $\forall \xi \in \Omega(G,k,\sigma_{x})$ it holds 
that
\begin{displaymath}
\mu\left(\xi |\sigma_{x}\right)= \mu\left((T\xi)|\eta_{x}\right)= \frac{1}{|\Omega_{\sigma}|}.
\end{displaymath}
This implies that 
$Pr[X^{\sigma}=\xi]=Pr[X^{\eta}=T\xi]$, $\forall \xi \in \Omega_{\sigma}.$
The lemma follows by noting that
\begin{displaymath}
\left (\sum_{\xi \in \Omega_{\sigma}} Pr[X^{\sigma}=\xi] \right )=1
\qquad \textrm{and} \qquad 
 \left (\sum_{\xi \in \Omega_{\sigma}} Pr[X^{\eta}=(T\xi)] \right )=1.
\end{displaymath}
\end{myproof}

\noindent
Let $\nu:[k]^{V}\times [k]^{V}\to [0,1]$ denote the joint distribution of the
colourings $X^{\sigma}$ and  $X^{\eta}$ in the coupling where $X^{\eta}=TX^{\sigma}$.
We close the section by providing a very useful property of $\nu$,
which we use in the disagreement percolation.

\begin{lemma}\label{lemma:worst-marginal}
For every $u \in V \backslash \{x\}$, let $N_{u}$ be the set that 
contains all the vertices which are adjacent to the vertex $u$ in $G $.
Also, let  ${\cal B}_u \subseteq [k]^{N_{u}}\times [k]^{N_{u}}$ be 
defined such that
\begin{displaymath}
{\cal B}_u=\{ \xi \in [k]^{N_{u}}\times [k]^{N_{u}}| \nu(\xi)> 0 \}.
\end{displaymath}
If $k>\Delta$, then it holds that 
\begin{displaymath}
\max_{\tau \in {\cal B}_u}
\nu(X^{\sigma}(u)\neq X^{\eta}(u)|\tau)
\leq \frac{1}{k-\Delta_u}
\end{displaymath}
where $\Delta_u$ is the degree of vertex $u$ in $G $.
\end{lemma}

\begin{myproof}
Let $G_X=(V_X,E_X)$, denote the induced subgraph of $G$ such that $v\in G_X$  if and only
if $X^{\sigma}(v)\neq X^{\eta}(v)$, in the coupling. We remind the reader that under both
$X^{\sigma}$ and $X^{\eta}$, $G_X$ is coloured using only the colours $\sigma_{x}$  and $\eta_{x}$.

There are two necessary conditions for some vertex $v \in V\backslash \{x\}$ to
be in $V_X$. The first one is that some vertex in $N_u$ should, also, belong to
$V_X$. This is due to the fact that $G_X$ is connected. The second is the following one: 
Assume that $w_1 \in N_u$ and $w_1\in V_X$. If there exists $w_2 \in  N_u \backslash \{w_1\}$ 
and $X^{\sigma}(w_2)\in \{\sigma_{x}, \eta_{x}\}$, then it should hold $X^{\sigma}(w_1)=X^{\sigma}(w_2)$.
This should hold under both $X^{\sigma}$ and $X^{\eta}$, $G_X$ is coloured using 
only the colours $\sigma_{x}$  and $\eta_{x}$.

Considering the two previous conditions the worst case of $X^{\sigma}(N_u)$ is the following:
At least one vertex in $N_{u}$ belongs to $V_X$, call this vertex $w$. No vertex in $N_{u}$ uses 
the colour $\{\sigma_{x}, \eta_{x}\}\backslash \{X^{\sigma}(w)\}$. $X^{\sigma}(N_u)$ is such that 
the number of different colour that are used is equal to $|N_u|$. In that case the probability of 
$u$ to belong to $V_X$ is  $\frac{1}{k-\Delta_u}$.  The lemma follows.
\end{myproof}

\noindent
Lemma \ref{lemma:worst-marginal} assumes that $k>\Delta$,  otherwise it holds
\begin{displaymath}
\max_{\tau \in {\cal B}_u}
\nu(X^{\sigma}(u)\neq X^{\eta}(u)|\tau)
\leq 1.
\end{displaymath}

\subsection{Proof of Theorem \ref{thrm:UNStructPercBnd}}
By  Theorem 1 and Corollary 1.1 in \cite{disagreement-percolation}, and Lemma \ref{lemma:worst-marginal} we get that
\begin{displaymath}
||\mu(\cdot|\sigma_{x})-\mu(\cdot|\eta_{x})||_{\Lambda}
\leq {\cal P}_{k,x }[\textrm{$\exists$ path of disagreement between $\{x\}$
and a vertex in $\Lambda$}].
\end{displaymath}

\noindent
We have to remark here that the coupling on which the disagreement percolation is based, has the following property: Let $t$ be the 
minimum integer such that there is no path of disagreement  connecting $x$ to $L(x, t)$. Then, our coupling specifies that no vertex in 
$L(x, t')$, for $t'\geq t$ can be disagreeing. This is a crucial  property of our coupling,  since otherwise we could not
apply the disagreement percolation technique (see \cite{haestrom}).

\section{Rest of the Proofs}\label{sec:RestProof}

\subsection{Proof of Proposition \ref{proposition:SchemaAccRel}}\label{sec:thrm:SchemaAccRel}
Let 
\begin{displaymath}
err_i={|Pr[X_{i}(v_i)\neq X_{i}(u_i)]-Pr[Y_{i}(v_i)\neq Y_{i}(u_i)]|} \qquad \textrm{for } 0\leq i\leq r-1.
\end{displaymath}
It holds that
\begin{displaymath}
\begin{array}{lcl}
\log{\cal Z} &=& \displaystyle 
\sum_{i=0}^{r-1}\log(P[Y_{i}(v_i)\neq Y_{i}(u_i)]) 
+ \log Z(G_0, k) 
\\ \vspace{-.3cm}\\
&\leq & \sum_{i=0}^{r-1}\log\left(
P[X_{i}(v_i)\neq X_{i}(u_i)]+err_i
\right)  + \log Z(G_0, k)
\\ \vspace{-.3cm}\\
&\leq & 
\sum_{i=0}^{r-1}\log\left(
P[X_{i}(v_i)\neq X_{i}(u_i)]\right) + \sum_{i=0}^{r-1}\log\left(1+\frac{err_i}{P[X_{i}(v_i)\neq X_{i}(u_i)]}\right)
+ \log Z(G_0, k) 
\\ \vspace{-.3cm}\\
&\leq & \displaystyle 
\log Z(G,k)+
\sum_{i=0}^{r-1}\log\left(1+\frac{err_i}{P[X_{i}(v_i)\neq X_{i}(u_i)]}\right)
\\ \vspace{-.3cm}\\
&\leq & \displaystyle \log Z(G,k)+
\sum_{i=0}^{r-1}\frac{err_i}{P[X_{i}(v_i)\neq X_{i}(u_i)]}.
\end{array}
\end{displaymath}
The final derivation follows by the fact that $\log(x)$ is an increasing
function (the base is of the logarithm is $e>1$) and by $1+x\leq e^x$, for any $x$.
Similarly we get the lower bound for $\log({\cal Z})$. The theorem follows.

\subsection{Proof of Proposition \ref{proposition:count-accuracy}}\label{sec:count-accuracy}
Proposition \ref{proposition:count-accuracy} follows as a corollary of the two following 
lemmas.
\begin{lemma}\label{lemma:Event2TVD}
It holds that
\begin{displaymath}
|Pr[X_i(v_i)\neq X_{i}(u_i)]-{P}r[Y_{i}(v_i)\neq Y_{i}(u_i)]| \leq 
\sum_{j=0}^{r_i-1}||\mu_{i,j}(\cdot)-\mu_{i,j+1}(\cdot)||_{\Psi_i}.
\end{displaymath}
\end{lemma}
\begin{myproof}
Let $\mu_{i,j}$ be the Gibbs distribution of the $k$-colourings of $G_{i,j}$.
It holds that
\begin{displaymath}
|Pr[X_i(v_i)\neq X_{i}(u_i)]-{P}r[X_{i,0}(v_i)\neq X_{i,0}(u_i)]| \leq 
\max_{A \subseteq [k]^{\Psi_i}}|\mu_{i,0}(A)-\mu_{i,r_i}(A)|
\leq ||\mu_{i,0}(\cdot)-\mu_{i,r_i}(\cdot)||_{\Psi_i}
\end{displaymath}
By the triangle inequality we get that
$||\mu_{i,0}(\cdot)-\mu_{i,r_i}(\cdot)||_{\Psi_i}\leq \sum_{j=0}^{r_i-1}
||\mu_{i,j}(\cdot)-\mu_{i,j+1}(\cdot)||_{\Psi_i}$
\end{myproof}

\begin{lemma}\label{lemma:simplfy-comparison}
Let $\Lambda$ be any subset of vertices of $G_{i,j}$ that does not
contain $v_{i,j}$ and $u_{i,j}$. It holds that 
\begin{displaymath}
||\mu_{i,j}(\cdot)-\mu_{i,j+1}(\cdot) ||_{\Lambda} \leq
C_{i,j} 
\max_{\sigma, \tau \in [k]^{\{v_{i,j}\}}}
\left\{ 
||\mu_{i,j}(\cdot|\sigma)-\mu_{i,j}(\cdot|\tau) ||_{\Lambda\cup\{u_{ij}\}}
+
||\mu_{i,j}(\cdot|\sigma)-\mu_{i,j}(\cdot|\tau) ||_{\{u_{ij}\}}
\right \}
\end{displaymath}
where $C_{ij}=C_{i,j}(G_{i,j}, k)=\max_{s,t \in [k]}\left \{(Pr[X_{i,j}(u_{i,j})=s| X_{i,j}(v_{i,j})=t])^{-2} \right \}$.
\end{lemma}
\begin{myproof}
Let $\Omega_{i,j}$ denote the set of $k$-colourings of $G_{ij}$ and $\mu_{ij}$ the uniform
distribution over $\Omega_{i,j}$. It is straightforward that
\begin{displaymath}
|| \mu_{i,j}(\cdot)- \mu_{i,j+1}(\cdot) ||_{\Lambda} \leq 
\max_{\sigma, \tau}||  \mu_{i,j}(\cdot|\sigma_{\Psi_{i,j}})-\mu_{i,j+1}(\cdot|\tau_{\Psi_{i,j}}) ||_{\Lambda},
\end{displaymath}
where $\tau$ varies in $\Omega_{i,j+1}$ and $\sigma$ varies in $\Omega_{i,j}$.
By the fact that $\Omega_{i,j+1} \subseteq \Omega_{i,j}$  and by the conditional independence,
it holds that $\mu_{i,j+1}(\cdot| \tau_{\Psi_{i,j}})=\mu_{i,j}(\cdot|\tau_{\Psi_{i,j}})$.
Hence, we have that
\begin{equation}\label{eq:Reduction2SameGraph}
||\mu_{i,j}(\cdot)-\mu_{i,j+1}(\cdot) ||_{\Lambda} \leq \max_{\sigma, \tau}
||\mu_{i,j}(\cdot|\sigma_{\Psi_{i,j}}) -\mu_{i,j}(\cdot|\tau_{\Psi_{i,j}})||_{\Lambda}.
\end{equation}
By definition (see (\ref{eq:TVDDefinition})), there exists a set 
${\cal A}\subseteq [k]^{\Lambda}$ such that

\begin{displaymath}
||\mu_{i,j}(\cdot|\sigma_{\Psi_{i,j}})-\mu_{i,j}(\cdot|\tau_{\Psi_{i,j}}) ||_{\Lambda} =
|\mu_{i,j}({\cal A}|\sigma_{\Psi_{i,j}})- \mu_{i,j}({\cal A}|\tau_{\Psi_{i,j}})|.
\end{displaymath}
Let $Q_{ij}=\mu_{ij}(\tau_{u_{ij}}|\tau_{v_{ij}})-\mu_{ij}(\sigma_{u_{ij}}|\sigma_{v_{ij}})$.
Using elementary probability theory relations we get the following:
\begin{displaymath}
\begin{array}{lcl}
|\mu_{i,j}({\cal A}|\sigma_{\Psi_{i,j}})- \mu_{i,j}({\cal A}|\tau_{\Psi_{i,j}})|
&\leq & \displaystyle \left |\frac{\mu_{i,j}(A,\tau_{u_{ij}}|\tau_{v_{ij}})}{\mu_{i,j}(\tau_{u_{ij}}|\tau_{v_{ij}})}-
 \frac{\mu_{i,j}(A,\sigma_{u_{ij}}|\sigma_{v_{ij}})}{\mu_{i,j}(\sigma_{u_{ij}}|\sigma_{v_{ij}})} \right |
\\ \vspace{-.3cm}\\
&\leq & \displaystyle \left |\frac{\mu_{i,j}(A,\tau_{u_{ij}}|\tau_{v_{ij}})}{\mu_{i,j}(\sigma_{u_{ij}}|\sigma_{v_{ij}})+Q_{ij}}-
 \frac{\mu_{i,j}(A,\sigma_{u_{ij}}|\sigma_{v_{ij}})}{\mu_{i,j}(\sigma_{u_{ij}}|\sigma_{v_{ij}})} \right|
\\ \vspace{-.3cm}\\
&\leq & \displaystyle \left |\frac{\mu_{i,j}(A,\tau_{u_{ij}}|\tau_{v_{ij}})}{\mu_{i,j}(\sigma_{u_{ij}}|\sigma_{v_{ij}})}-
 \frac{\mu_{i,j}(A,\sigma_{u_{ij}}|\sigma_{v_{ij}})}{\mu_{i,j}(\sigma_{u_{ij}}|\sigma_{v_{ij}})} \right|+
\\ \vspace{-.3cm}\\
&&\displaystyle +\frac{|Q_{i,j}|}{\mu_{i,j}(\tau_{u_{ij}}|\tau_{v_{ij}})\mu_{i,j}(\sigma_{u_{ij}}|\sigma_{v_{ij}})}.
\end{array}
\end{displaymath}
It is direct to see that
\begin{displaymath}
\begin{array}{c}
\displaystyle 
|\mu_{i,j}(A,\tau_{u_{ij}}|\tau_{v_{ij}})-\mu_{i,j}(A,\sigma_{u_{ij}}|\sigma_{v_{ij}})|\leq 
\max_{\tau, \sigma}||\mu_{i,j}(\cdot|\tau_{v_{ij}})- \mu_{i,j}(\cdot|\sigma_{v_{ij}})||_{\Lambda^*}
\\ \vspace{-.3cm}\\
\displaystyle 
|\mu_{ij}(\tau_{u_{ij}}|\tau_{v_{ij}})-\mu_{ij}(\sigma_{u_{ij}}|\sigma_{v_{ij}})|\leq 
\max_{\tau, \sigma}||\mu_{i,j}(\cdot|\tau_{v_{ij}})- \mu_{i,j}(\cdot|\sigma_{v_{ij}})||_{u_{ij}},
\end{array}
\end{displaymath}
where $\Lambda^*=\Lambda\cup\{u_{ij}\}$. The lemma follows.
\end{myproof}

\subsection{Proof of Lemma \ref{lemma:Count2Sample}}\label{sec:lemma:Count2Sample}
Consider the sequence of subgraphs ${\cal G}(G)=G_0, \ldots, G_r$, 
where $r=|E|$ and $G_0$ is empty. 
Consider, also, the following telescopic relation
\begin{displaymath}
|\Omega(G,k)|=|\Omega(G_0,k)|\cdot \prod_{i=0}^{r-1}\frac{|\Omega(G_{i+1}, k)|}
{|\Omega(G_{i}, k)|}=k^n\cdot \prod_{i=0}^{r-1}\frac{|\Omega(G_{i+1}, k)|}
{|\Omega(G_{i}, k)|}.
\end{displaymath}
The lemma will follow by showing that
\begin{displaymath}
Pr[X_i(u_i)\neq X_i(v_i)]=\frac{|\Omega(G_{i+1}, k)|} {|\Omega(G_{i}, k)|}.
\end{displaymath}
The above relation clearly holds by noting the following:
The set of $k$-colourings of $G_{i+1}$  is the same as the subset 
of $k$-colourings of $G_{i}$ that contains all the colourings that 
assign $v_{i}$ and $u_{i}$ different colours. The lemma follows.

\subsection{Proof of Lemma \ref{lemma:reconstruction}.}\label{sec:reconstruction}
\begin{displaymath}
\begin{array}{lcl}
||\mu(\cdot|\sigma_x)-\mu(\cdot)||_{\Lambda}
&=& \displaystyle 
\frac{1}{2}\sum_{\sigma_{\Lambda}\in [k]^{\Lambda}}
|\mu(\sigma_{\Lambda}|\sigma_x)-\mu(\sigma_{\Lambda})|
\\ 
& = & \displaystyle 
\frac{k}{2}\mu(\sigma_x)\sum_{\sigma_{\Lambda}\in [k]^{\Lambda}}
|\mu(\sigma_{\Lambda}|\sigma_x)-\mu(\sigma_{\Lambda})|
\\ 
& = & \displaystyle 
\frac{k}{2}\sum_{\sigma_{\Lambda}\in [k]^{\Lambda}}
\mu(\sigma_{\Lambda})|\mu(\sigma_x|\sigma_{\Lambda})-\mu(\sigma_x)|
\\ 
& \leq & \displaystyle 
\frac{k}{2}\sum_{\sigma_{\Lambda}\in [k]^{\Lambda}}
\mu(\sigma_{\Lambda})\sum_{\tau_x \in [k]}|\mu(\tau_x|\sigma_{\Lambda})-\mu(\tau_x)|
\\ 
& \leq & \displaystyle 
k \sum_{\sigma_{\Lambda}\in [k]^{\Lambda}}
\mu(\sigma_{\Lambda})||\mu(\cdot |\sigma_{\Lambda})-\mu(\cdot) ||_x.
\end{array}
\end{displaymath}
Noting that it holds
$$||\mu(\cdot|\sigma_{x})- \mu(\cdot|\tau_{x})||_{\Lambda}\leq
||\mu(\cdot|\sigma_{x})- \mu(\cdot)||_{\Lambda}+
||\mu(\cdot)- \mu(\cdot|\tau_{x})||_{\Lambda},
$$
the lemma follows.
\remove{
For the first inequality we have the following:
\begin{displaymath}
\begin{array}{lcl}
\displaystyle \sum_{A \in [k]^{\Lambda}}\mu(A)\cdot ||\mu(\cdot| A)-\mu(\cdot)||_x 
&=& \displaystyle \frac{1}{2}\cdot \sum_{A \in [k]^{\Lambda}}\sum_{\xi_x\in [k]^x} \mu(A) \cdot |\mu(\xi_x| A)-\mu(\xi_x)| \\
&=& \displaystyle \frac{1}{2}\cdot \sum_{\xi_x\in [k]^x} \sum_{A \in [k]^{\Lambda}} \mu(\xi_x) \cdot |\mu(A|\xi_x)-\mu(A)| \\
&=& \displaystyle \sum_{\xi_x\in [k]^x} \mu(\xi_x) \cdot \frac{1}{2}\left( \sum_{A \in [k]^{\Lambda}}|\mu(A|\xi_x)-\mu(A)| \right)\\
&=& \displaystyle \sum_{\xi_x\in [k]^x} \mu(\xi_x) \cdot ||\mu(\cdot|\xi_x)-\mu(\cdot)||_{\Lambda}\\
&\leq& \displaystyle \sum_{\xi_x\in [k]^x} \mu(\xi_x) \cdot \max_{\sigma_x, \tau_x \in [k]^x}||\mu(\cdot|\sigma_x)-\mu(\cdot|\tau_x) ||_{\Lambda} \\ \vspace{-.3cm}\\
&=& \max_{\sigma_x, \tau_x \in [k]^x}||\mu(\cdot|\sigma_x)-\mu(\cdot|\tau_x) ||_{\Lambda} \\
\end{array}
\end{displaymath}
}

\subsection{Proof of Lemma \ref{lemma:REdges}}\label{sec:lemma:REdges}
Let $\epsilon=1/(10\log(d))$.  Assume that after removing all the edges in $R$ 
there are two cycles of length at most $\epsilon \log n$ which are connected, i.e.
these two cycles share edges. Then, there  must exist a subgraph of $G_{n,d/n}$ 
that contains at  most $2\epsilon\log n$ vertices  while the number of edges 
exceeds  by 1, or more,  the number of vertices.

Let $D$ be the event that in $G_{n,d/n}$ there exists a set of $r$ vertices which have $r+1$ edges between them. For $r \leq \epsilon \log n$ we have the following:
\begin{displaymath}
\begin{array}{lcl}
\displaystyle Pr[D] & \leq & \displaystyle \sum_{r=1}^{\epsilon\log n} {n \choose r} { {r \choose 2} \choose r+1} (d/n)^{r+1} (1-d/n)^{{r \choose 2}-(r+1)}
\\ \vspace{-.3cm} \\
& \leq & \displaystyle \sum_{r=1}^{\epsilon \log n} \left( \frac{n e}{r} \right )^r \left ( \frac{r^2 e}{2(r+1)} \right )^{r+1} (d/n)^{r+1} 
\leq 
\frac{e \cdot d}{2n}\sum_{r=1}^{\epsilon \log n} \left ( \frac{e^2 d}{2} \right )^{r}
\\ \vspace{-.3cm}\\
& \leq & 
\displaystyle \frac{C}{n} \left (\frac{e^2d}{2} \right)^{\epsilon \log n}.
\end{array}
\end{displaymath}
Having $\epsilon \cdot \log(e^2d/2)< 1 $, the quantity in the r.h.s. of the last inequality is $o(1)$, 
in particular it is of order $\Theta(n^{\epsilon \log (e^2d/2)-1})$. Thus, for $\epsilon =1/(10\log(d))$ 
there is no connected component that contains two cycles with probability at least $1-n^{-0.85}$.

Let $C_l$ denote the number of cycles of length at most $l$ in $G(n,d/n)$. It is direct to
show that $E[C_l]\leq 2d^l$. Furthermore, 
$E[C_{\epsilon\log n}]\leq 2n^{1/10}$.
It is not hard to see that the expected number of edges whose one end is on a cycle of length less than
$\epsilon \log n$ is $O(n^{1/10}\log^2 n)$. That is $E[|R|]= O(n^{1/10} \log^2 n)$.

Employing the Markov inequality, we have $Pr[|R|\geq n^{3/10}] =O(n^{-0.2}/\log^2 n)$
while $Pr[C_{\epsilon\log n}\geq n^{3/10}]\leq 2n^{-0.2}$. The lemma follows.

\vspace*{1cm}

\noindent
{\bf Acknowledgement.}
The author would like to thank the anonymous reviewer as well as  Prof. Mike Paterson for the time they
spent to read this manuscript and for their corrections and their suggestions to improve the presentation
of the result. Also, the author would like  to thank Amin Coja-Oghlan for the numerous fruitful discussions.

\end{document}